\documentclass[twocolumn]{aastex631}
\usepackage{amsmath}
\usepackage{amssymb}
\usepackage{float}
\shorttitle{Planetary Phase equilibria}
\shortauthors{Young et al. }

\newcounter{rxn}

\begin{document}

\title{Phase Equilibria of Sub-Neptunes and Super-Earths}

\correspondingauthor{Edward D. Young}
\email{eyoung@epss.ucla.edu}

\author{Edward D. Young}
\affiliation{Department of Earth, Planetary, and Space Sciences\\
University of California, Los Angeles\\
Los Angeles, CA 90095, USA}

\author{Lars Stixrude}
\affiliation{Department of Earth, Planetary, and Space Sciences\\
University of California, Los Angeles\\
Los Angeles, CA 90095, USA}

\author{James G. Rogers}
\affiliation{Department of Earth, Planetary, and Space Sciences\\
University of California, Los Angeles\\
Los Angeles, CA 90095, USA}

\author{Hilke E. Schlichting}
\affiliation{Department of Earth, Planetary, and Space Sciences\\
University of California, Los Angeles\\
Los Angeles, CA 90095, USA}

\author{Sarah P. Marcum}
\affiliation{Department of Earth, Planetary, and Space Sciences\\
University of California, Los Angeles\\
Los Angeles, CA 90095, USA}

\begin{abstract}
We investigate the consequences of non-ideal chemical interaction between silicate and overlying hydrogen-rich envelopes for rocky planets using basic tenets of phase equilibria.  Based on our current understanding of the temperature and pressure conditions for complete miscibility of silicate and hydrogen, we find that the silicate-hydrogen binary solvus will dictate the nature of atmospheres and internal layering in rocky planets that garnered H$_2$-rich primary atmospheres.   The temperatures at the surfaces of supercritical magma oceans will correspond to the silicate-hydrogen solvus.  As a result, the radial positions of supercritical magma ocean-atmosphere interfaces, rather than their temperatures and pressures, should reflect the thermal states of these planets. The conditions prescribed by the solvus influence the structure of the atmosphere, and thus the transit radii of sub-Neptunes.   Separation of iron-rich metal to form metal cores in sub-Neptunes and super-Earths is not assured due to prospects for neutral buoyancy of metal in silicate melt induced by dissolution of H, Si, and O in the metal at high temperatures.  
\end{abstract}

\keywords{Exoplanet structure, Exoplanet atmospheric composition, Exoplanet evolution}

%\begin{linenumbers}

\section{Introduction}
Super-Earths and sub-Neptunes are the most abundant extrasolar planets based on our current census of the Galaxy \citep[e.g.][]{fressin2013a}. Super-Earths constitute a population of rocky exoplanets of roughly Earth-like compositions more massive than Earth, while sub-Neptunes constitute a population of larger planets likely also with rock/metal cores (we use the term ``core" here when referring to the combination of the silicate mantle and metal core, and ``metal core" when referring to the iron-rich phase only) but surrounded by significant H/He envelopes comprising a few percent of the total mass of the planets \citep{gupta2019a,rogers2020b}. It is generally considered that sub-Neptunes retained their primary atmospheres while the less massive super-Earths also formed with H$_2$-rich atmospheres but lost them over time \citep{owen2013a,gupta2020a}. A revelation from transit surveys of these planets is, therefore, that hydrogen-rich atmospheres were apparently integral to their formation and evolution, raising the spectre of the ubiquity of hydrogen-rich atmospheres during rocky planet formation in the Galaxy \citep[e.g.][]{Bean2021}. Although there are no representatives of this class of planets in our Solar System, our planets may be the products of variations on the same theme.
Neptune and Uranus could have formed by similar accretion processes, but involving more volatile-rich ``cores" \citep{Helled2020a}, while Earth may be an example of a planet formed from embryos with relatively meager, but still important, hydrogen-rich atmospheres \citep{Young2023a}.  The picture that emerges is one in which the various types of rocky planets form a continuum related by the processes of atmospheric loss and atmosphere-interior interactions that vary with mass and access to hydrogen.  

 The rocky planets and their precursor embryos are evidently born hot as a result of the conversion of gravitational potential energy into thermal energy (in addition to other sources of heat such as short-lived radionuclides) during accretion. As a result, bodies sufficiently hot to form magma oceans will remain hot for up to billions of years (Gyr) if blanketed by an optically thick H$_2$-rich atmospheres \citep[e.g.][]{Misener2022}. Chemical equilibrium between the condensed ``cores" (i.e., silicate and metal) and overlying atmospheres contributes higher molecular weight species to the primary atmospheres \citep{Kite2019, Kite_2020b, Schlichting_Young_2022}, exacerbating the potential for retaining heat.  Interaction between molten silicate, molten metal, and hydrogen-rich atmospheres is therefore a crucial aspect of sub-giant planet formation and evolution. 
 
 In this work we consider the temperature and pressure conditions attending the formation of planets with substantial primary atmospheres, and examine the implications of the relevant phase equilibria at those conditions for the structure of those planets. For this purpose we employ the simplest of thermodynamic relationships between silicate magma oceans and overlying hydrogen-rich atmospheres, that of a non-ideal binary mixture.

 The paper is structured as follows.  In section \ref{section:TandP} we describe the temperatures and pressures of sub-Neptune cores and atmospheres in the absence of chemical exchange between the reservoirs.  In section \ref{section:reactions} we describe reactions between silicate magma oceans and hydrogen atmospheres in terms of their miscibility. Section \ref{section:implications} describes the implications of complete miscibility of silicate and hydrogen for sub-Neptune cores and atmospheres, and section \ref{section:structure} presents a general structure for sub-Neptunes based on the miscibility of silicate and hydrogen.  Conclusions are summarized in section \ref{section:conclusions}.

\section{Temperatures and Pressures at Magma Ocean Surfaces}
\label{section:TandP}
 
Temperatures and pressures at atmosphere-magma ocean interfaces early in the life of a planet can comprise extreme conditions that go beyond the bounds of our present understanding of the physical chemistry of silicate melts, oxide melts, and impure atmospheres.  In some cases, these phases are likely completely miscible, for example, defying our concept of a planetary surface.  We are accustomed to these circumstances for the interiors of giant planets in relatively simple chemical systems (e.g., supercritical hydrogen), but the implications for rocky planets or planets whose mass budget is dominated by 'solids' rather than hydrogen (i.e. Sub-Neptunes, Uranus and Neptune) are just beginning to be explored.

 In models meant to constrain the mass-radius relationships in sub-Neptunes, the focus is correctly placed on the temperture and pressure structures of the hydrogen-rich envelopes.  As a result, the temperatures and pressures at the base of hydrogen-rich envelopes are well studied \citep[e.g.,][]{Chen2016,ginzburg2016a}. As an approximation, the magma ocean ``cores" are often considered to be isothermal because differences in temperature of order  $\sim 2 \times$, as imposed by adiabatic temperature-pressure profiles \citep{Solomatov2009,Chachan2018,Young2019}, for example, are not especially important for the overall thermal budget \citep[e.g.,][]{Fortney2007,Lopez2014,ginzburg2016a,Brouwers2018}.  Given the relatively loose constraints from planet formation models, the initial conditions at the surfaces of the magma oceans beneath H$_2$-rich envelopes are often treated as free parameters \citep[e.g.,][]{Chen2016,Vazan2018,Chachan2018, Modirrousta-Galian2023}. However, in the present work, factors of order $2 \times$ are critical to interpretations of the phase equilibria of the core-envelope interface, and we therefore seek  tighter constraints on the likely temperatures and pressures in the planet's interior. The issue is what are the conditions that satisfy both the potential temperatures of vigorously convecting magma oceans and the temperatures and pressures imposed by the overlying atmospheres. 

In this section, we offer some simple calculations to illustrate the ranges of temperatures and pressures relevant for atmosphere-magma ocean interfaces. To do so, we constrain the relevant temperatures and pressures for the phase equilibria with a reprisal of the basic features of planet accretion with H$_2$-rich atmospheres. The key result is that accretion of the hydrogen-rich primary atmospheres evidenced by many exoplanets with masses of $\sim 2$ to 10 $M_\oplus$ results in temperatures at the base of the atmospheres of several thousands of degrees at GPa pressures.    We show that this is a robust conclusion \citep[e.g.,][]{ginzburg2016a} that can be illustrated with the salient features of different models for  planets that accrete atmospheres from their host protoplanetary disks.    The same conclusion can also be inferred from fitting mass-radius relations for planets with rocky cores and hydrogen dominated envelopes to the observed Sub-Neptune exoplanet population \citep[e.g.,][]{lopez2013a,chen2016a}.  

\subsection{Gas accretion}

 We envision a rocky core of mass $M_c$ and radius $R_c$ accreting rapidly compared to the disk life time. The core captures gas from the disk forming an atmosphere of mass $M$.  This captured atmosphere initially has luminosity sufficiently large that heat can be transported through the atmosphere only by convection so that the temperature profile is adiabatic.  
 For a convecting, isentropic gas $d\ln{T}/d\ln{P}=R/{\rm C}_{\rm P} $, where $R$ is the gas constant and $\rm C_P$ is the molar isobaric heat capcity. Treating the hydrogen atmosphere as an ideal gas composed of H$_2$ molecules, $R/{\rm C}_{\rm P}=\gamma/(\gamma-1) $ where $\gamma=7/5$ is the ratio of isobaric to isochoric heat capacity.  For hydrostatic pressure as a function of radial position $r$, $d\ln{P}/dr=-\mu g/(k_{\rm B}T)$, where $\mu$ is the mass of a gas molecule, $k_{\rm B}$ is the Boltzmann constant, $g=GM_c/r^2$ is the acceleration due to gravity, and $G$ is the universal gravitational constant. We have, therefore, a temperature profile given by $dT/dr=((1-\gamma)/\gamma) \mu g/k_{\rm B}$.  We assume the atmosphere blends into the disk at the Bondi radius, so it proves convenient to recast the constants in terms of the Bondi radius,

 \begin{equation}
R_{\rm B}=\frac{2GM_c\mu}{\gamma k_{\rm B}T_{\rm eq}}.
\label{eq:bondi}
\end{equation}
 
  \noindent such that upon integration we have   
 \begin{equation}
 \frac{T(r)}{T_{\rm eq}} =1 + \frac{R_{\rm B}^{\prime}}{r} - \frac{R_{\rm B}^{\prime}}{R_{\rm B}},
 \label{eq:adiabaticT}
 \end{equation}

 \noindent In Equation \ref{eq:adiabaticT}, $T_{\rm eq}$ is the temperature of the disk and    
\begin{equation}
    R_{\rm B}^{\prime}=\frac{\gamma-1}{2}R_{\rm B}.
\end{equation}

In the regime of interest, $R_{\rm B}^{\prime}>>R_c$ so that the temperature at the surface of the magma ocean $T_c=T(R_c)$ is largely independent of the properties of the surrounding disk, in which case
\begin{equation}
    T_c(\mathrm{K})=\frac{\gamma - 1}{\gamma}\frac{G M_c \mu}{k_{\rm B} R_c}=5050\left( \frac{M_c}{M_{\oplus}} \right)^{3/4},
\label{eq:Tc}
\end{equation}
where we have assumed that the $R_c \propto M_c^{\beta}$ with $\beta=1/4$ \citep{valencia2006}.  
%Similarly, density and pressure of the atmosphere at the core surface in the limit $R_{\rm B}^{\prime}>>R_c$
%\begin{equation}
%    \rho_c=5050 \left( \frac{\rho_d}{10^{-6} \mathrm{g/cm}^3}\right) \left( \frac{M_c}{M_{\oplus}} \right)^{\frac{3}{4(\gamma-1)}}
%\end{equation}
%\begin{equation}
%xx
%\end{equation}

This analysis shows that the base of the atmosphere is expected to have temperatures far exceeding the melting point of silicates.  For example, for $M_c=4M_{\oplus}$, $T_c=14300$ K, as compared with the melting temperature of silicates of $\sim 2000$ K.  An initially molten surface is also likely from the point of view of rocky core accretion, as shown below.

\subsection{Adiabatic core accretion}
\label{section:Adiabatic_core_accretion}
The maximal temperature of the core as a whole is obtained by assuming adiabatic conversion of gravitational potential energy into heat during accretion \citep[e.g.,][]{Treves1988}, i.e., with no radiative losses, and recognizing that latent heats of melting are negligible.  In this case we have $T = (3/5) GM_{\rm c}/(R_{\rm p}c)$ where $c$ is the specific heat.  For a fiducial planet mass of 4$M_\oplus$ and a typical silicate melt $c$ of $1500$ J/(kg K) (e.g., from ${\rm C_p} \sim 4nR$ with $n=5$ atoms per formula unit and $0.1$ kg/mole for melt, \citealt{DeKoker2009}),  $T = 70700$ K. 
%More generally, 
%\begin{equation}
%    \Delta T (\mathrm{K}) = 38,000 \left( \frac{M_c}{M_{\oplus}} \right) ^{3/4} \frac{c}{1000 \mathrm{ ~J/kg/K}}
%\end{equation}
%These values are likely to be overestimates as some of the gravitational energy is lost tduring accretion. Nevertheless, it %shows that surface temperatures as high as 14,000 K are reasonable. 
A body with this mass-integrated temperature would have an isentropic potential temperature (i.e., magma ocean surface temperature if blanketed by a dense atmosphere) of $\sim 50,000$ K based on the equations for melt adiabats used here.  However, in order to retain an H$_2$-rich  atmosphere comprising 10\% by mass of the body, or for that matter an atmosphere of just a few percent hydrogen, with an equilibrium temperature of about 300 K (a relatively low value to be conservative), reasonable values for hydrogen-rich sub-Neptunes, a maximum temperature of $\sim 14,000$ K at the base of the atmosphere is allowed \citep{ginzburg2016a}. The surface temperature of the magma ocean as calculated in the adiabatic accretion scenario is therefore far greater than the maximum surface temperature for retention of an H$_2$-rich atmosphere. Here it is the atmosphere that controls the temperature structure of the core by setting the $T$ and $P$ boundary conditions for the magma ocean once the core is cool enough to bind the atmosphere.

\subsection{Diabatic core accretion}
An alternative method for estimating the temperature of the surface of the core might be to calculate the luminosity due to non-adiabatic (i.e., diabatic) accretion.  Here a protracted accretion  of planetesimals and/or pebbles is spread over some accretion time interval with quasi-continuous radiative losses.  The luminosity for protoplanets in the absence of atmospheres is then $L_{\rm acc}=GM_{\rm c} \dot{m}/R_{\rm p}$ \citep{Benfield1950,Frank2002,rafikov2006a}. With this scenario, the surface black-body radiation temperature is $T_{\rm B} = (L_{\rm acc}/(\sigma 4 \pi R^2_{\rm p}))^{1/4}$, yielding a value of $T_{\rm B}$=830 K for a 4$M_\oplus$ planet with an accretion rate of $\dot{m} = 1M_{\oplus}/{\rm Myr}$. Ten times higher accretion rates, i.e. $\dot{m} = 10M_{\oplus}/{\rm Myr}$,  required for the most massive sub-Neptune exoplanets to ensure formation before the gas disk dissipates, yield values for $T_{\rm B}$ of about 1660 K.
However, once these bodies approach Mars size, they will accrete primary atmospheres, and their internal temperature and luminosity will be controlled by radiation through their atmospheres. Once an optically thick atmosphere accretes, diffusion across the radiative-convective boundary yields a luminosity of

\begin{equation}\label{eqn:ginzburg15}
L_{\rm atm}=\frac{64\pi}{3}\frac{\sigma T_{\rm rcb}^4 R_{\rm B}((\gamma -1)/\gamma)}{\kappa\  \rho_{\rm rcb}},
\end{equation}
where $\kappa$ and $\rho_{\rm rcb}$ are the mean opacity and density at the radiative-convective boundary, respectively \citep{ginzburg2016a}.  
%For a $0.5M_\oplus$ protoplanet, the maximum mass fraction of the primary atmosphere is about 1\%, and the ratio $L_{\rm atm}/L_{\rm acc}$ is $0.6$.  The timescale required to dissipate the entirety of the gravitational potential energy for this body $E_{\rm grav} = (3/5) G(0.5M_\oplus)^2/(0.84R_\oplus)$,  is $E_{\rm grav}/L_{\rm atm}=1\times 10^6$ yrs. 
For a $1M_\oplus$ protoplanet, with a 2\% by mass atmosphere, we find that the ratio of $L_{\rm atm}/L_{\rm acc}$ is $0.11$. The timescale required to dissipate the entirety of the gravitational potential energy for this body $E_{\rm grav} = (3/5) G(1M_\oplus)^2/(1R_\oplus)$,  is $E_{\rm grav}/L_{\rm atm}=6\times 10^6$ yrs.  Although comparable in magnitude, this timescale is longer than typical observed gas disk lifetimes. Therefore, core temperatures of typical protoplanets forming with a primordial atmosphere should exceed the temperature for the atmosphere-free case, $T_{\rm B}$, by at least a factor of several, since their interiors can't radiate away all of their internal energy between successive accretion events.  Here again, the magma ocean surface temperatures are expected to be at least several thousand degrees.

\subsection{Cooling}
As the atmosphere cools, the luminosity $L$ diminishes, and the atmosphere develops an upper zone in which heat is transported by photons.    From Eddington's equation for radiative diffusion,   the temperature gradient in this radiative zone, $\nabla_{\rm rad} = dT/dr$, is
\begin{equation}
    \nabla_{\rm rad}=-\frac{3 \kappa \rho L}{64 \pi \sigma T^3 r^2},
    \label{eq_nablarad}
\end{equation}
where $\kappa$ is the opacity, $\rho$ is the density, and $\sigma$ is the Stefan-Boltzmann constant.  For the opacity $\kappa=\kappa(P,T)$, we adopt the parameterization of \cite{freedman2014a}.  For the density, we assume hydrostatic equilibrium:
\begin{equation}
    \frac{d \ln \rho}{d r}= -\frac{G M_c \mu}{k_{\rm B} T r^2},
\end{equation}
and for the pressure, we adopt the ideal gas equation of state:
\begin{equation}
    P=\frac{\rho}{\mu} k_{\rm B}T.
\label{eq:idealgas}
\end{equation}
At any radius in the atmosphere, we assume that the temperature gradient is
\begin{equation}
    \frac{dT}{dr}=\frac{\nabla_{\mathrm{rad}} \nabla_{\mathrm{conv}}}{\nabla_{\mathrm{rad}}+\nabla_{\mathrm{conv}}},
    \label{eq_effectivemass}
\end{equation}
where the convective temperature gradient is adiabatic such that
\begin{equation}
    \nabla_{\mathrm{conv}}=-\frac{\gamma-1}{\gamma} \frac{GM_c \mu}{k_{\rm B} r^2}.
    \label{eq_nablaconv}
\end{equation}
Equations \ref{eq_nablarad}-\ref{eq_nablaconv} complete our formulation of atmospheric structure.

In our approach, the radiative-convective boundary does not appear explicitly, as it does in many previous studies, which assume an abrupt switch from radiative to convective heat transport at the radiative convective boundary $R_{\mathrm{rcb}}$ \citep[e.g.,][]{ginzburg2016a, owen2017a}.  Instead, heat transport switches over gradually from radiative to convective with increasing density according to Eq.~\ref{eq_effectivemass}.  We derive the radiative-convective boundary from our atmospheric structure solutions, taking $R_{\mathrm{rcb}}$ to be that radius $r$ at which $\nabla_{\mathrm{rad}}(r)=\nabla_{\mathrm{conv}}(r)$.  

As an illustration, we integrate Eqs. \ref{eq_nablarad}-\ref{eq_nablaconv} with boundary conditions suitable for many observed sub-Neptune exoplanets with $M_c=4M_{\oplus}$ and $T_{\rm eq}=T(r=R_{\rm B})$=1000 K, $\rho_{\mathrm{disk}}=\rho(r=R_{\rm B})$=10$^{-6}$ g/cm$^3$, corresponding to an orbital radius of 0.08 AU (orbital period 8 days) in a minimum mass solar nebula \citep{Hayashi_1981}.  
\begin{figure}
   \includegraphics[width=0.45\textwidth]{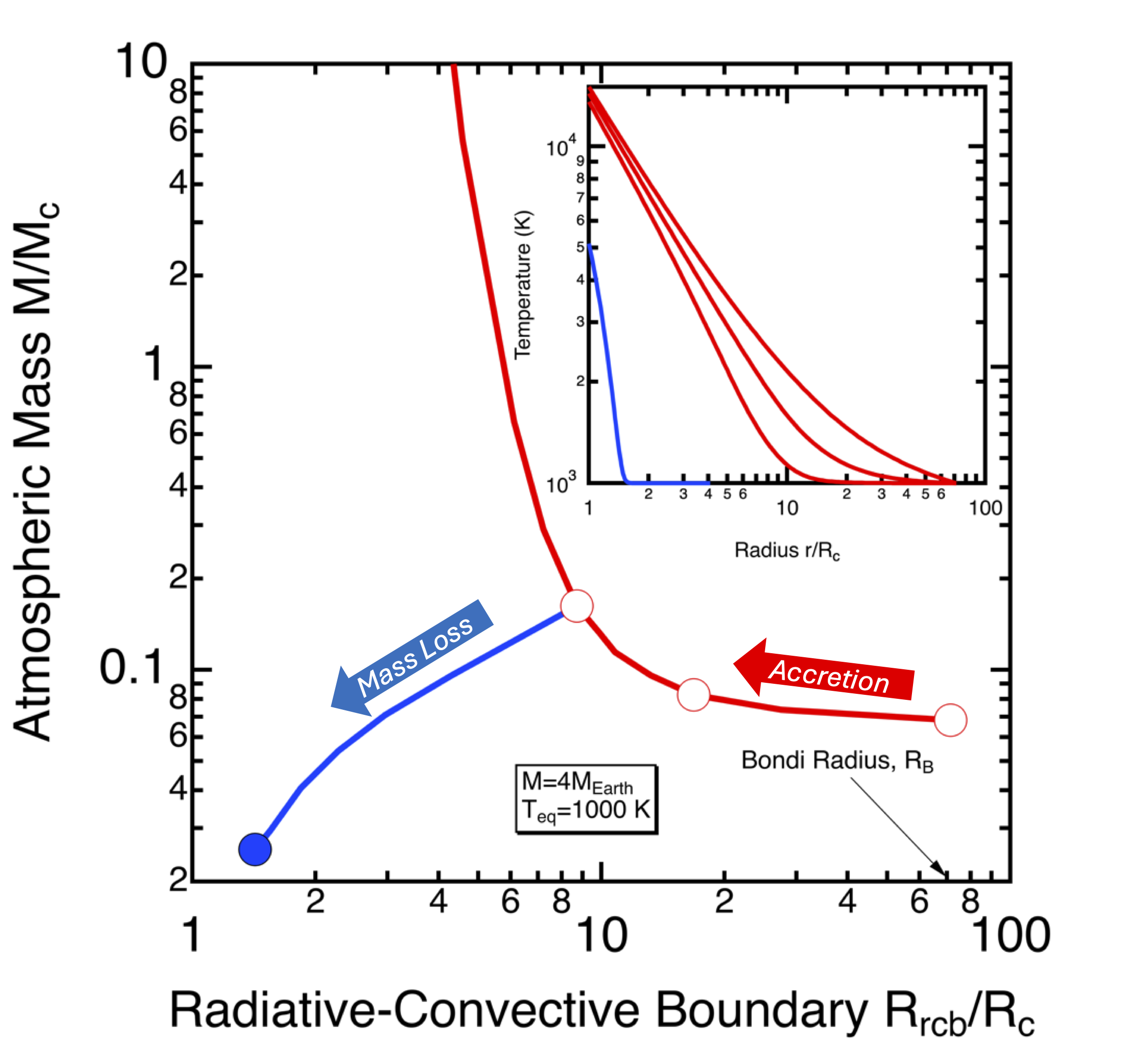}
    \caption{Evolution of the structure of the atmosphere during the accretion (red) and mass loss (blue) phases of planetary evolution, with arrows indicating the direction of evolution.  (Main figure) Atmospheric mass in units of $M_c$ plotted against the radiative-convective boundary $R_{\mathrm{rcb}}$ in units of $R_c$.  (Inset) Temperature structure of the atmosphere at different points during the evolution: the three red curves correspond to the conditions at the three open circles in the main figure with the leftmost curve being the most evolved, the blue curve corresponds to the blue circle in the main figure.  }
\label{fig_structure}
\end{figure}
Decreasing luminosity yields a family of atmospheric structures upon cooling (Figure~\ref{fig_structure}).  Initially the atmospheric structure is adiabatic, with no radiative zone.  As the atmosphere cools and the luminosity diminishes the radiative zone grows and $R_{\mathrm{rcb}}$ shrinks, producing a nearly isothermal outer region.  As the planet cools, the atmosphere grows in mass $M$ by accretion from the surrounding disk (Figure \ref{fig_structure}).  The phenomenon of accretion on cooling has been examined in previous studies \citep{lee2015a, ginzburg2016a}.  

\subsection{Mass loss following disk dispersal}
Atmospheres are  susceptible to erosion by photoevaporation for planets proximal to their host stars \citep{owen2017a, owen2019a}, core-powered mass loss \citep{ginzburg2016a}, and ``spontaneous-mass loss"/``boil off" once the confining protoplanetary disk dissipates \citep{owen2016a,ginzburg2016a,Rogers2023}. In the case of some sub-Neptunes, atmospheres can be stable for of order $10^9$ years \citep{ginzburg2018a,rogers2020b, Markham2022}. On the other hand, in other circumstances (e.g., with favorable disk dispersal timescales compared to planetary accretion and cooling timescales) significant fractions of the primary atmospheres (e.g., $>90$\%) can be lost in $10^6$ years during disk dispersal \citep{ginzburg2016a,Rogers2023}.  Here we consider a simple representation of atmospheric mass loss with implications for the temperature structure of the planet. 

Following dispersal of the protoplanetary disk, the planet's primary atmosphere is no longer supported by the pressure of the surrounding disk and the planet loses atmospheric mass on a time scale
\begin{equation}
    t_{\mathrm{loss}}=M/\dot{M}.
\end{equation}
As mass is lost, the planet continues to cool on a time scale
\begin{equation}
    t_{\mathrm{cool}}=E/\dot{E}.
\end{equation}
The atmospheric mass is effectively
\begin{equation}
    M=\int_{R_c}^{R_{\rm B}} 4 \pi \rho (r)  r^2 dr 
\end{equation}
while the total energy of atmosphere and core is
\begin{equation}
E=E_{\mathrm{atm}} + E_c, 
\end{equation}
where 
\begin{equation}
E_{\mathrm{atm}} = \int_{R_c}^{R_{\rm B}} \left( -\frac{G M_c}{r} + \frac{1}{\gamma - 1} \frac{k_{\rm B} T}{\mu} \right) 4 \pi \rho r^2 dr 
\end{equation}
and
\begin{equation}
    E_c=\int_0^{R_c} c T 4 \pi \rho r^2 dr
    \label{eq:ecore}
\end{equation}
represent the energy of the atmosphere and the core, respectively. 

We calculate the evolutionary history of our planet following disk dispersal by assuming $t_{\mathrm{cool}} = t_{\mathrm{loss}}$, an equality found to hold approximately in previous studies \citep{Misener2021}.  In this case thermal evolution and mass loss are linked such that the fractional loss of mass in any time interval is equal to the fractional loss of energy: $M/M_d=E/E_d$, where subscript $d$ refers to the values at the time of disk dispersal.  For our initial condition along this trajectory, corresponding to the time of disk dispersal, we take our atmospheric structure computed above at $M=0.16 M_c$, similar to one of the cases studied by \cite{Misener2021}.  
The trajectory of the planet in this phase is very different from the trajectory that it follows while the disk is present (Figure~\ref{fig_structure}).  Atmospheric mass and $R_{\rm rcb}$ shrink together. The atmosphere is stable against mass loss: $E_{\mathrm{atm}}<0$, i.e. the gravitational potential energy of the atmosphere exceeds in magnitude the thermal energy of the atmosphere.

\subsection{Temperature and pressure at the magma ocean surface with time}
\label{section:time_evolution}

\begin{figure}
   \includegraphics[width=0.45\textwidth]{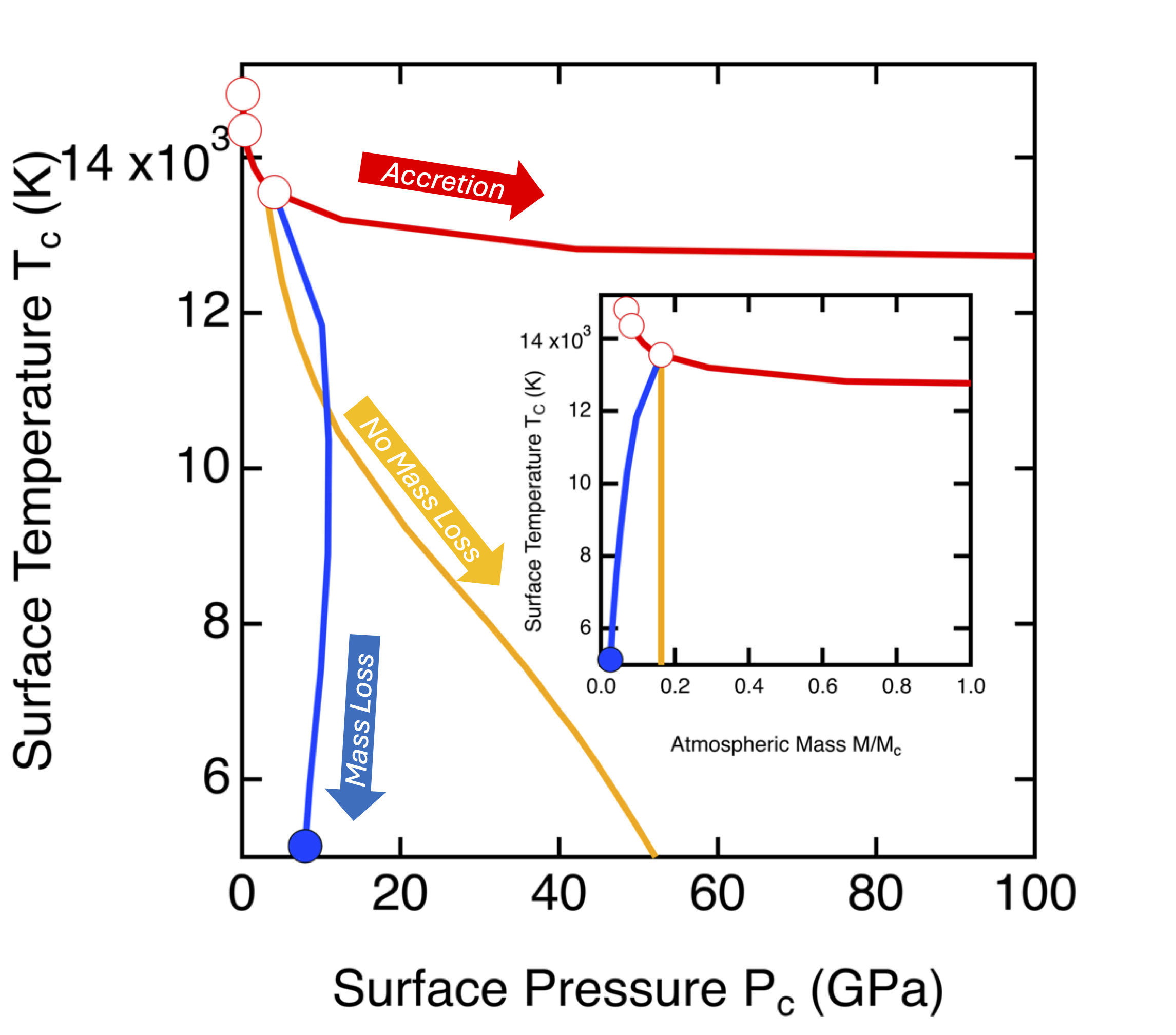}
    \caption{Evolution of the conditions at the surface of the core during the accretion phase (red) and the mass loss phase (blue solid).  Surface temperature versus surface pressure is shown in the main figure and surface temperature versus atmospheric mass is shown in the inset.    Also shown is a cooling path without mass loss (green).    The symbols in the main figure and the inset correspond to those states highlighted by symbols in Figure~\ref{fig_structure}.}
\label{fig_coresurface}
\end{figure}
 
We find that the pressure and temperature at the core surface follows two distinct trends  (Figure~\ref{fig_coresurface}).  During accretion, the core surface follows a nearly isothermal path in which pressure grows with atmospheric mass by a factor of 200 up to $M=2M_c/3$, while the temperature decreases by only 20 \% as the outer radiative zone grows.  Following disk dispersal, the core surface follows a  nearly isobaric path, with temperature decreasing by a factor of 3, while the pressure varies non-monotonically by a factor of 2.  The reason the pressure changes little with mass loss is that $R_{\mathrm{\rm rcb}}$ diminishes during mass loss (Figure ~\ref{fig_structure}), concentrating a greater proportion of the atmospheric mass close to the core surface.   For comparison, we also show in Figure \ref{fig_coresurface} the case for cooling with no mass loss.  In this scenario, pressures are greater at the magma ocean surface as again $R_{\mathrm{\rm rcb}}$ decreases (closer to the condensed surface), but with no compensating loss of mass.     

The temperature-pressure structure of the planet core (silicate magma ocean and metal core) can be constructed by matching the potential temperature of the adiabatic magma ocean to the basal atmospheric $T$. 
For the core, we assume a layered structure, consisting of a rocky mantle and an Fe-rich metallic core in relative mass proportion equal to that of Earth (2:1).  We assume that the interior is isentropic with surface conditions matching those at the base of the atmosphere at $r=R_c$: $T=T_c$, $P=P_c$.  To derive the core structure, we solve the system of equations \citep[e.g.,][]{seager2007a}:
\begin{equation}
    \frac{dm}{dr}=4 \pi r^2 \rho,
\label{eq:dmdr}
\end{equation}
\begin{equation}
    \frac{dP}{dr}=-\frac{Gm \rho}{r^2},
\label{eq:dPdr}
\end{equation}
and,
\begin{equation}
\frac{dT}{dr}=\frac{\Gamma T \rho g}{K_S},
\label{eq:dTdr}
\end{equation}
where $m$ is the mass contained within radius $r$, $\Gamma$ is the Gr\"uneisen parameter and $K_S$ is the adiabatic bulk modulus, together with the equations of state of mantle and core material (see Appendix).
Numerically integrating Equations (\ref{eq:dmdr}) through (\ref{eq:dTdr}) for core and mantle yields a density and temperature profile for the core (see Appendix). 

\begin{figure}
   \includegraphics[width=0.45\textwidth]{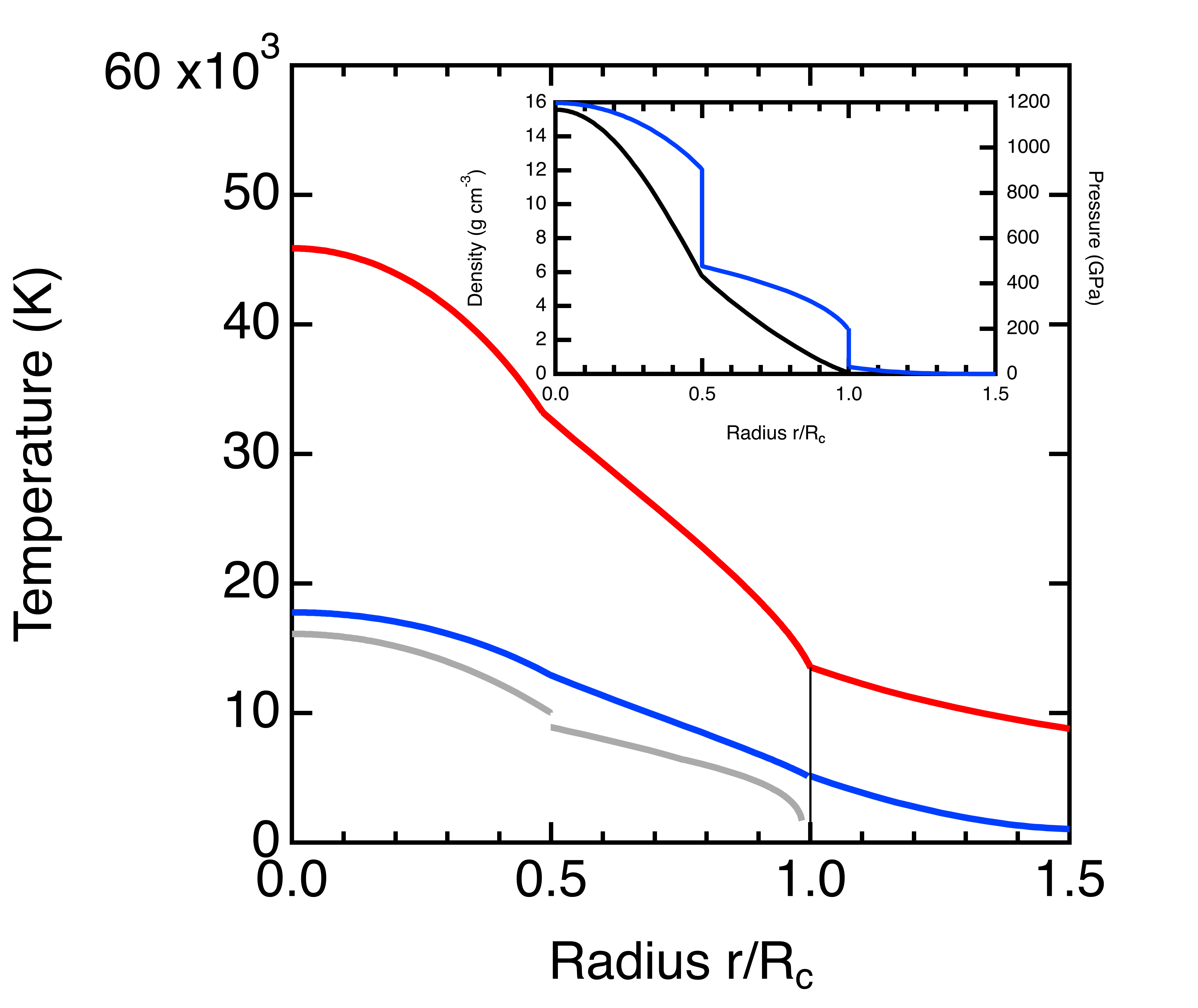}
    \caption{Temperature of the planet at disk dispersal (red) and after mass loss (blue) compared with estimates of the melting temperature of silicates and iron-rich alloy (grey lines) shown over the domains of rock (0.5$<r/R_c<$1.0) and metal ($r/R_c<$0.5), respectively.  The red and blue curves correspond to the states highlighted by symbols in Figure~\ref{fig_structure} at disk dispersal ($M=0.16M_c$) and after mass loss ($M=0.025M_c$).  The vertical line shows the surface of the core.  The inset shows the density (blue) and pressure (black, right-hand axis) of the planet following mass loss.  }
\label{fig_core}
\end{figure}

We use Equations (\ref{eq:dmdr}) through (\ref{eq:dTdr}) to calculate the core pressures and temperatures in equilibrium with the atmospheres with time (see Appendix).
The core in our example maintains very high temperatures through time, reaching 45,000 K at the center immediately following disk dispersal, cooling to 18,000 K at the center following atmospheric mass loss (Figure \ref{fig_core}).  The temperature throughout the core exceeds the melting temperature; the core is a magma ocean that is completely molten from center to surface.  The pressure reaches 1200 GPa at the center, more than 3 times that at Earth's center, and the density reaches 16 g cm$^{-3}$, 20 \% greater than at Earth's center.  

 Because the $dT/dP$-slopes of silicate melt adiabats are shallower than those of their liquidi and solidi \citep[e.g.,][]{Stixrude2014}, cooling will lead to progressive solidification of the magma oceans from the interior outward, locking in the results of phase equilibria incurred at higher temperatures.

\section{Silicate-H$_2$ Miscibility}
\label{section:reactions}

The simplified examples of atmosphere and core accretion described above would suggest that hydrogen-rich atmospheres and magma oceans are in contact during much of the evolution of sub-Neptunes, with magma ocean surface temperatures of $\sim 5000$ K to $> 10000$ K, and pressures of order $\sim 1$ to 50 GPa (e.g., Figure \ref{fig_coresurface}). While a useful guide to relevant conditions, the results outlined above are predicated on simple physical systems with no chemical interaction. What is more, they assume immiscible molten rock and atmospheres regardless of $T$ and $P$.  Consider, for example, that the meaning of surface temperature or pressure as described above  where the core and atmosphere become completely miscible is lost at these conditions. The complexities of these chemical systems are the subject of a number of recent studies \citep[e.g.,][]{Kite2019, Kite_2020b, Schlichting_Young_2022,Markham2022,Misener2023,Rogers2024}. In particular, \cite{Markham2022} and \cite{Misener2023} considered feedbacks between the chemical interactions between atmospheres and magma oceans and the density structure of the planet's atmospheres and their capacity for convection. Most relevant to the present study, \cite{Markham2022} highlighted the fact that the silicate-hydrogen system might approach supercriticality at conditions expected to occur at atmosphere-magma ocean interfaces, here after referred to as AMOIs.  The structure of the planets in question changes markedly where miscibility obtains at conditions that would otherwise define the AMOIs.

\begin{figure}[h]
\centering
   \includegraphics[width=0.45\textwidth]{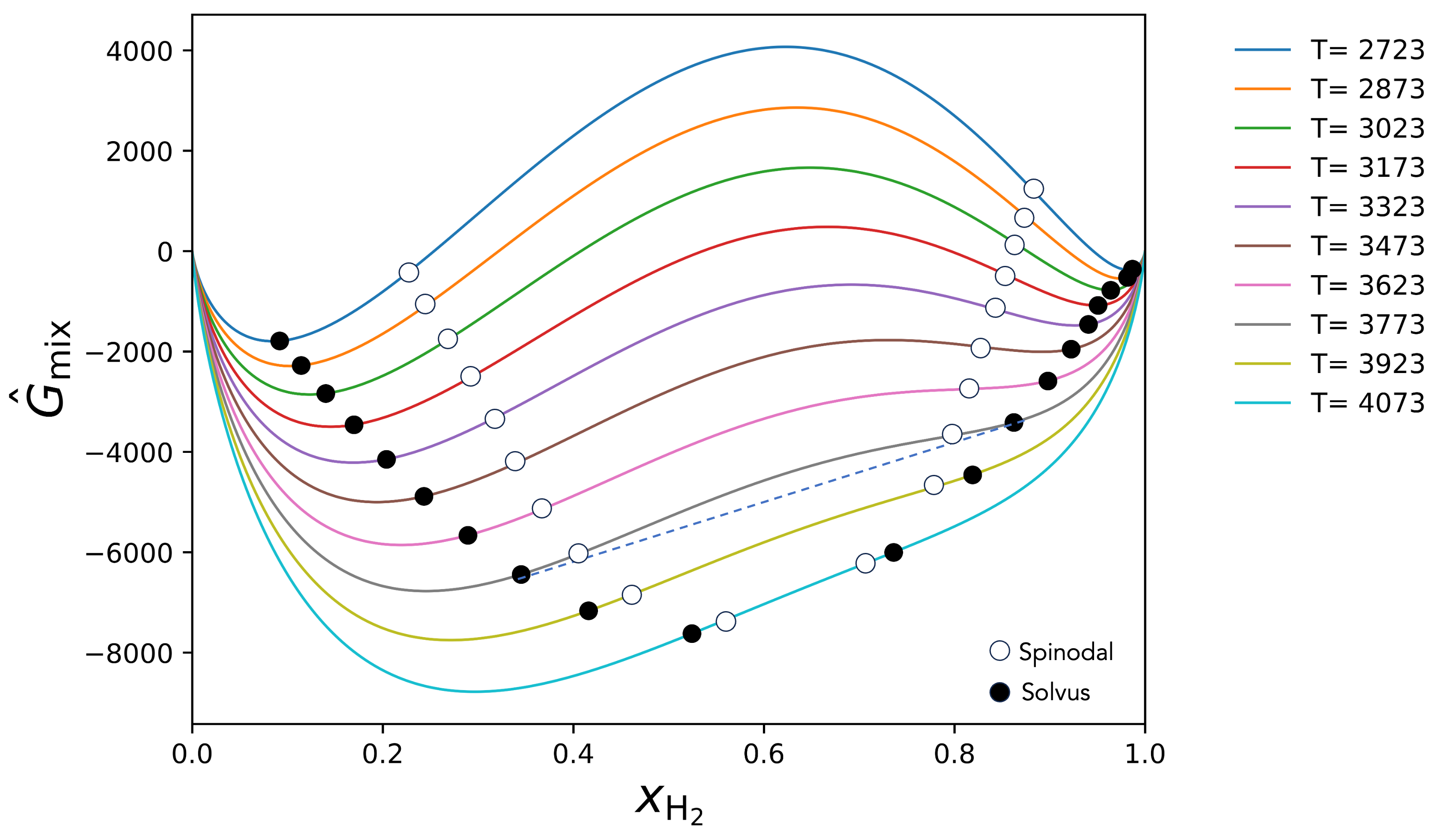}
    \caption{ The free energy of mixing as a function of temperature and the mole fraction of hydrogen  for a binary silicate-H$_2$ system based on an asymmetrical mixing model and a consolute temperature near 4,000 K.  The black points demarcate the coexisting equilibrium compositions of silicate (left side) and atmosphere (right side) for different temperatures that define the solvus curve in Figure \ref{fig:solvus1percent}. The dashed tanget line for the 3,773 K isotherm illustrates the derivation of coexisting phase compositions. The open points define the spinodal curve.  A single phase with a composition between the spinodal points spontaneously decomposes into two phases with the metastable compositions indicated by the spinodes.  The curves were produced using $W_{\rm A} = 130000$ J/mole,  $W_{\rm B} = -50000$, and $\tau_{S} = 10100$.    }
\label{fig:Gsurface}
\end{figure}

Mapping the $T$ and $P$-dependent silicate-hydrogen miscibility would move us toward  a more thorough and unified understanding of planetary evolution, from Earths to Neptune-like planets. Two-phase ab-initio calculations for mixing between silicates and H$_2$ \citep{Gilmore2023} suggest that indeed, the conditions predicted to occur at AMOIs appear to broadly coincide with silicate-H$_2$ miscibility.   To facilitate communication across disciplines, we will use the phase equilibria term ``solvus" for the coexistence curve separating the regions of two-phase and single-phase thermodynamic stability for silicate and hydrogen. The boundary, more generally referred to as the binodal, could alternatively be described with a spinodal curve if it demarcates the conditions for spontaneous decomposition of one phase into two phases. The spinodal refers to the boundary for spontaneity of the phase transition rather than thermodynamic stability (Figure \ref{fig:Gsurface}) \citep[e.g.,][]{Allabar2018}.  \cite{Gilmore2023} provide a detailed analysis of the system MgSiO$_3$-H$_2$ using density functional theory molecular dynamics (DFT-MD) simulations, including relevant thermodynamic parameters.  Here we are concerned with the implications of the mere existence of the sovlus at conditions relevant for hydrogen-rich envelopes and hydrogen-rich silicate melts. For this purpose, it is sufficient to consider the results for a bulk hydrogen composition of about 4\% by mass, indicating that the silicate-H$_2$ system is completely miscible above about 4,000 K at $3.5$ GPa while two phases exist at  3,000 K at similar pressures.  These results suggest that the implied $T$ and $P$ for the solvus for this bulk composition are near those expected for a sub-Neptune with order a few weight per cent H$_2$. \begin{new} We note that we are exploring the consequences of complete miscibility between the hydrogen-rich gaseous phase and silicate melt.  Our emphasis is not the precise temperature and pressure of the solvus (binodal), and our analysis is generally robust with respect to uncertainties in these details. \end{new}

The implications for silicate-hydrogen miscibility can be portrayed using simple binary phase diagrams. Such diagrams could be constructed assuming symmetrical, non-ideal mixing (on a molar basis), conforming to a regular solution in which the interaction parameter $W$ is related to the consolute temperature, $T_{\rm  c}$ (peak $T$ of the symmetrical mixing solvus) by $W=2RT_{\rm c}$.  The molar change in enthalpy associated with mixing, referred to as the excess enthalpy, is then $\Delta \hat{H}_{\rm EX}=x_{\rm sil}x_{\rm H_2}W$, where $x_i$ refers to the mole fraction of species $i$.  
The change in the Gibbs free energy due to mixing, $\Delta \hat{G}_{\rm mix}$, is computed from this non-ideal enthalpy of mixing and an ideal entropy of mixing, yielding
\begin{multline}
\Delta \hat{G}_{\rm mix}=x_{\rm sil}x_{\rm H_2}W + \\
RT(x_{\rm H_2} \ln{x_{\rm H_2}}+(1-x_{\rm H_2}) \ln{(1-x_{\rm H_2}))},
\label{eqn:Gmixing}
\end{multline}
where the entropy of mixing is $\Delta \hat{S}_{\rm mix}=-R(x_{\rm H_2} \ln{x_{\rm H_2}}+(1-x_{\rm H_2}) \ln{(1-x_{\rm H_2})})$.  The Gibbs free energy of mixing for the $W$ implied by a consolute temperature of $4000$ K is 66500 J/mole.  
  The entropy of mixing here is based on mixing of the two molecular moieties.  It is a place holder for a more realistic formulation for the configurational entropy that involves speciation in the melt.

By analogy with the water-H$_2$ \citep{Gupta2024_water} and albite-water  \citep{Mibe2007,Makhluf2016} systems, the more general case would include the possibility that the non-ideal mixing is asymmetrical, reflecting greater solubility of the simpler molecular species in melt as opposed to the reverse.  In this case, a sub-regular solution model \citep{Hardy1953}, or a quasi-regular solution model can be used to describe the binary system. The excess enthalpy of mixing is in this case defined by two interaction parameters, where $\Delta \hat{H}_{\rm EX} = (W_{\rm A}+ W_{\rm B}x_{\rm sil})x_{\rm H_2}x_{\rm sil}$. The non-ideal mixing can then be described by the excess free energy, $\Delta \hat{G}_{\rm EX}$ of the form \citep{Onel2016}

\begin{equation}
\Delta \hat{G}_{\rm EX} = (W_{\rm A}+ W_{\rm B}x_{\rm sil})x_{\rm H_2}x_{\rm sil}(1-T/\tau_{{\rm S}}),
\label{eqn:Gxs}
\end{equation}
where $\tau_{\rm S}$ accounts for the temperature dependence of the excess free energy term, and the excess entropy of mixing is by definition $\Delta \hat{S}_{\rm EX} = (W_{\rm A}+ W_{\rm B}x_{\rm sil})x_{\rm H_2}x_{\rm sil}/\tau_{\rm S}$.
The free energy of mixing is, in the asymmetrical case, 

\begin{multline}
\Delta \hat{G}_{\rm mix}=(W_{\rm A}+ W_{\rm B}x_{\rm sil})x_{\rm H_2}x_{\rm sil}(1-T/\tau{{\rm S}})+ \\
RT(x_{\rm H_2} \ln{x_{\rm H_2}}+(1-x_{\rm H_2}) \ln{(1-x_{\rm H_2}))}.
\label{eqn:G_subregular}
\end{multline}

\noindent Figure \ref{fig:Gsurface} shows an example of the asymmetrical free energy of mixing as a function of temperature and the mole fraction of H$_2$ that satisfies the inference that the crest of the solvus may lie towards hydrogen, again by analogy with the albite-water system where the crest of the solvus is further from the silicate melt in composition, and that the crest of the solvus is at about 4,000 K. The temperature-dependent equilibrium compositions in this binary system, specified by the mole fractions of hydrogen comprising the silicate and H$_2$-rich atmosphere phases, respectively, are obtained by the tangents to the free energy surfaces at each temperature in Figure \ref{fig:Gsurface}. 

\begin{figure}
   \centering
   \includegraphics[width=0.45\textwidth]{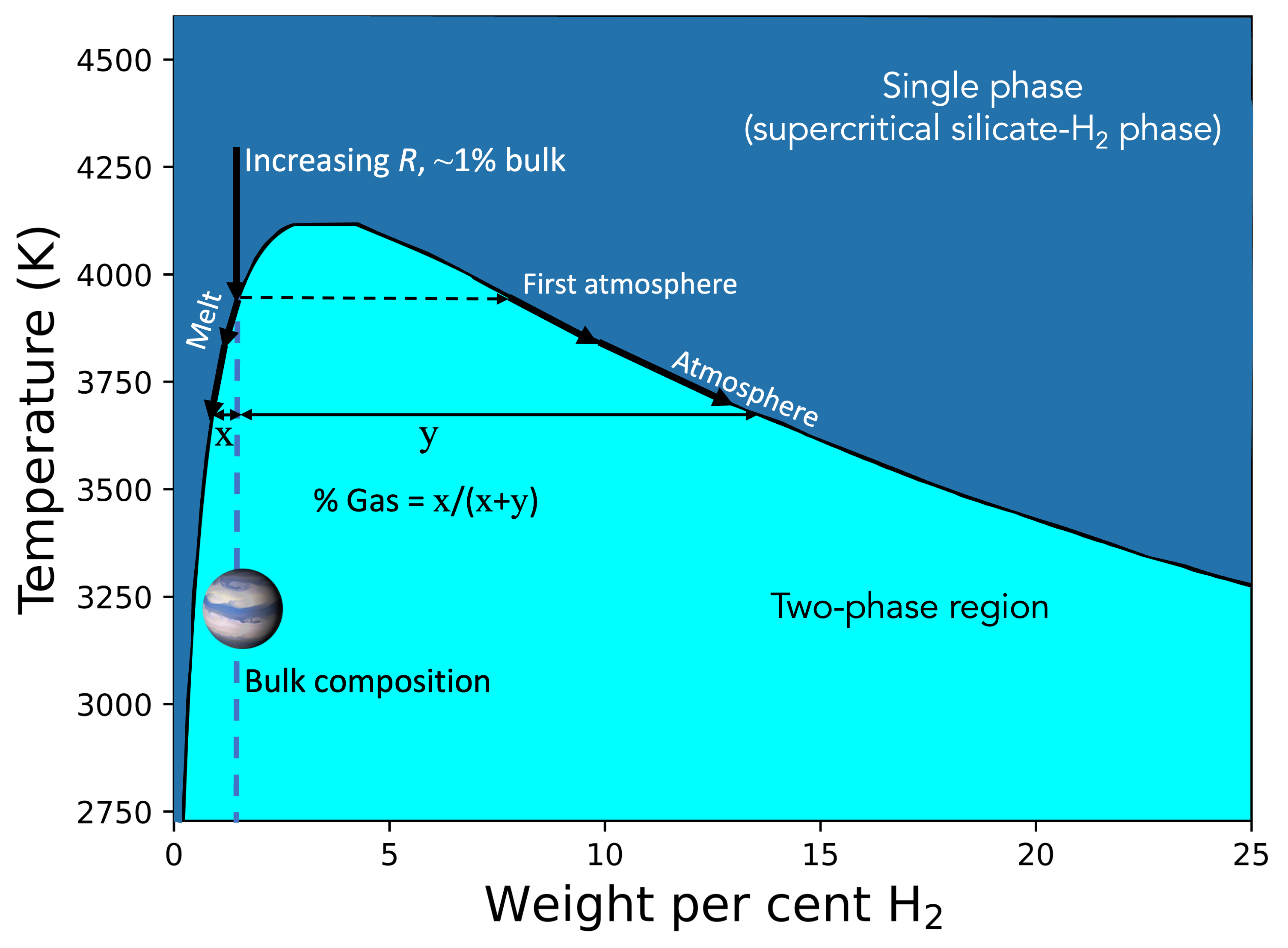}
    \caption{  A simple subregular solution solvus for a binary silicate-H$_2$ system based on the free energy surface shown in Figure \ref{fig:Gsurface}. Compositions of coexisting phases along the solvus curve are represented by points at equal temperature.  Representative coexisting phase compositions are shown by the horizontal lines connecting the two compositions.  The lever rule can be applied to obtain the mass fraction of each phase (using line segments x and y as shown).  Cooling paths with increasing distance from planet center ($R/R_{\rm p}$) for 1\% H$_2$ by mass is shown for illustration by the arrows. Planets symbolize relevant regions of bulk composition.  (Planet image credits: Pablo Carlos Budassi; NASA).    }
\label{fig:solvus1percent}
\end{figure}

The isobaric phase diagram derived from the free energy surfaces in Figure \ref{fig:Gsurface} is shown in Figure \ref{fig:solvus1percent} where the composition axis has been converted to weight per cent hydrogen (assuming MgSiO$_3$ for the formula unit of silicate) for comparison with the convention of describing planet compositions. In Figure \ref{fig:solvus1percent}, the concentrations of H$_2$ that define the compositions of two coexisting phases are obtained from two points on the solvus curve at a given temperature.  The lever rule can be used to specify the relative mass fractions of the two phases represented by the points.   For example, the mass fraction of exsolving gas (``atmosphere") at  3,700 K    is obtained from the length of line segment $x$ divided by the total distance between the two points on the solvus at that temperature, $x + y$. 

 \begin{figure}
   \centering
   \includegraphics[width=0.45\textwidth]{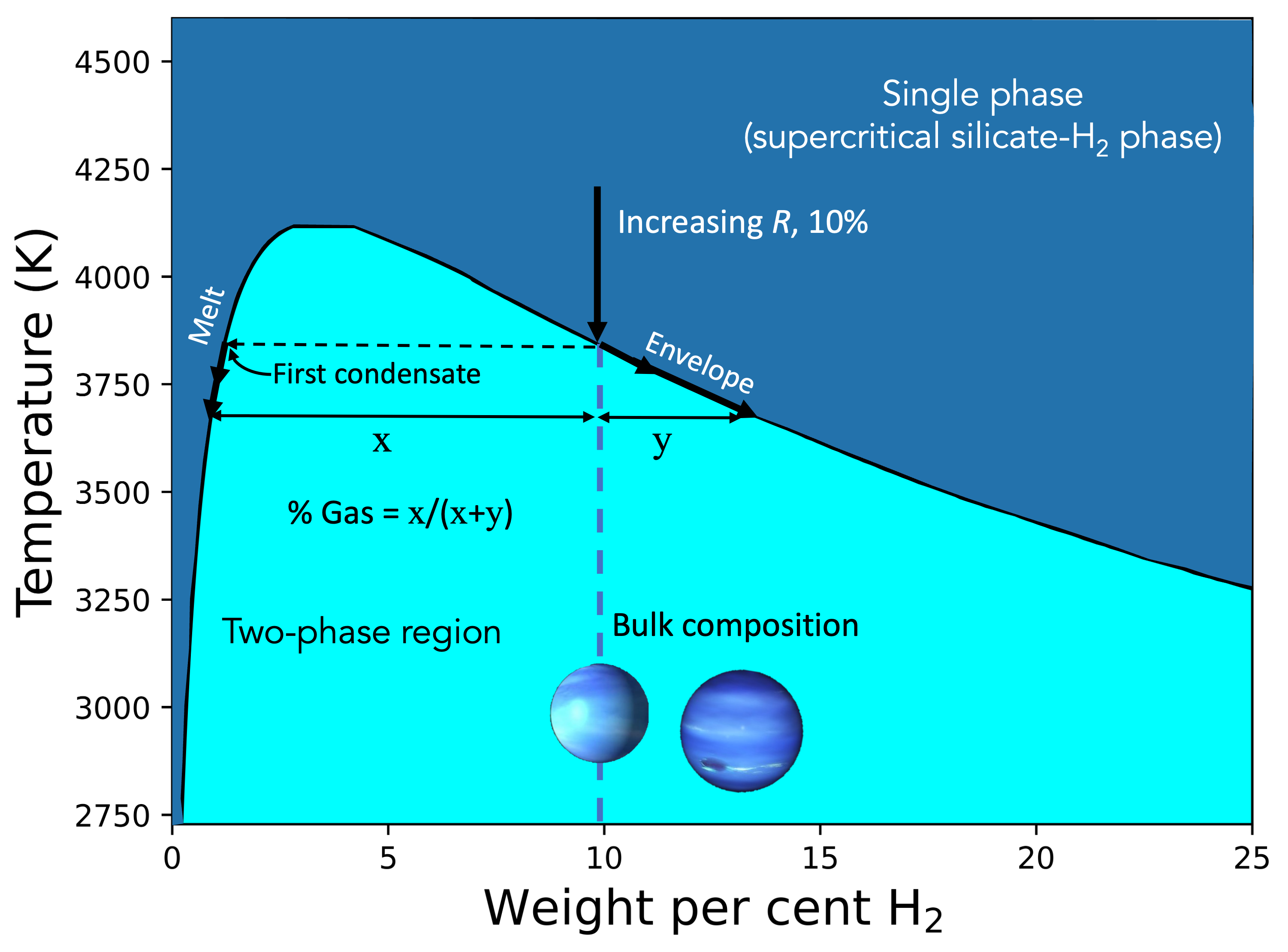}
    \caption{  Same as Figure \ref{fig:solvus1percent} but with paths with increasing distance from planet center ($R/R_{\rm p}$) for 10\% H$_2$ by mass shown by arrows. Planets symbolize relevant regions of bulk composition.  (Planet image credits: Pablo Carlos Budassi; NASA).   }
\label{fig:solvus10percent}
\end{figure}

   The crest of this hypothetical, asymmetrical solvus occurs at about 4\% by mass H$_2$. The crest of the solvus would be at about 2\% H$_2$ in the case of symmetrical mixing. Planets, or regions in planets, with $\le 4$\% hydrogen are completely miscible mixtures of silicate and hydrogen at temperatures above $\sim$ 4,100 K in this example.    At these temperatures, there is no discrete core and atmosphere and the concept of solubility of H$_2$ in a magma ocean, for example, does not apply.  Taken literally, and ignoring pressure effects for the moment, with cooling outward through the planet, a distinct atmosphere exsolves with a composition that is more hydrogen rich than the bulk (First atmosphere, Figure \ref{fig:solvus1percent}), but still of relatively high molecular weight due to the large fraction of silicate present.  This breach of the solvus defines the surface of the planet. If the surface cools,  more exsolution of atmosphere occurs. 
 The molecular weight of the atmosphere decreases as the mass fraction of atmosphere increases. 
 
   Conversely, for   supercritical magma oceans    with $> 4$\% hydrogen by mass, cooling below $\sim 4,100$ K (the precise $T$ of the solvus depends on the bulk composition), results in rainout of silicate from the slicate-hydrogen mixture as the temperature of the solvus at that composition is attained (Figure \ref{fig:solvus10percent}).  Here the mass of the envelope is dominated by the more hydrogen-rich phase, with silicates remaining subordinate in mass until temperatures are significantly below the solvus.  For 10\% H$_2$ bulk composition case (Figure \ref{fig:solvus10percent}), silicate first exsolves at a temperature of about 3,800 K but the mass of silicate is subordinate to the mass of H$_2$ until $T$ is $<$ 3,300 K.    Continuous removal of the silicate by settling to the core would result in distillation of the envelope to more and more hydrogen rich compositions.   

\begin{figure}
\centering
   \includegraphics[width=0.46\textwidth]{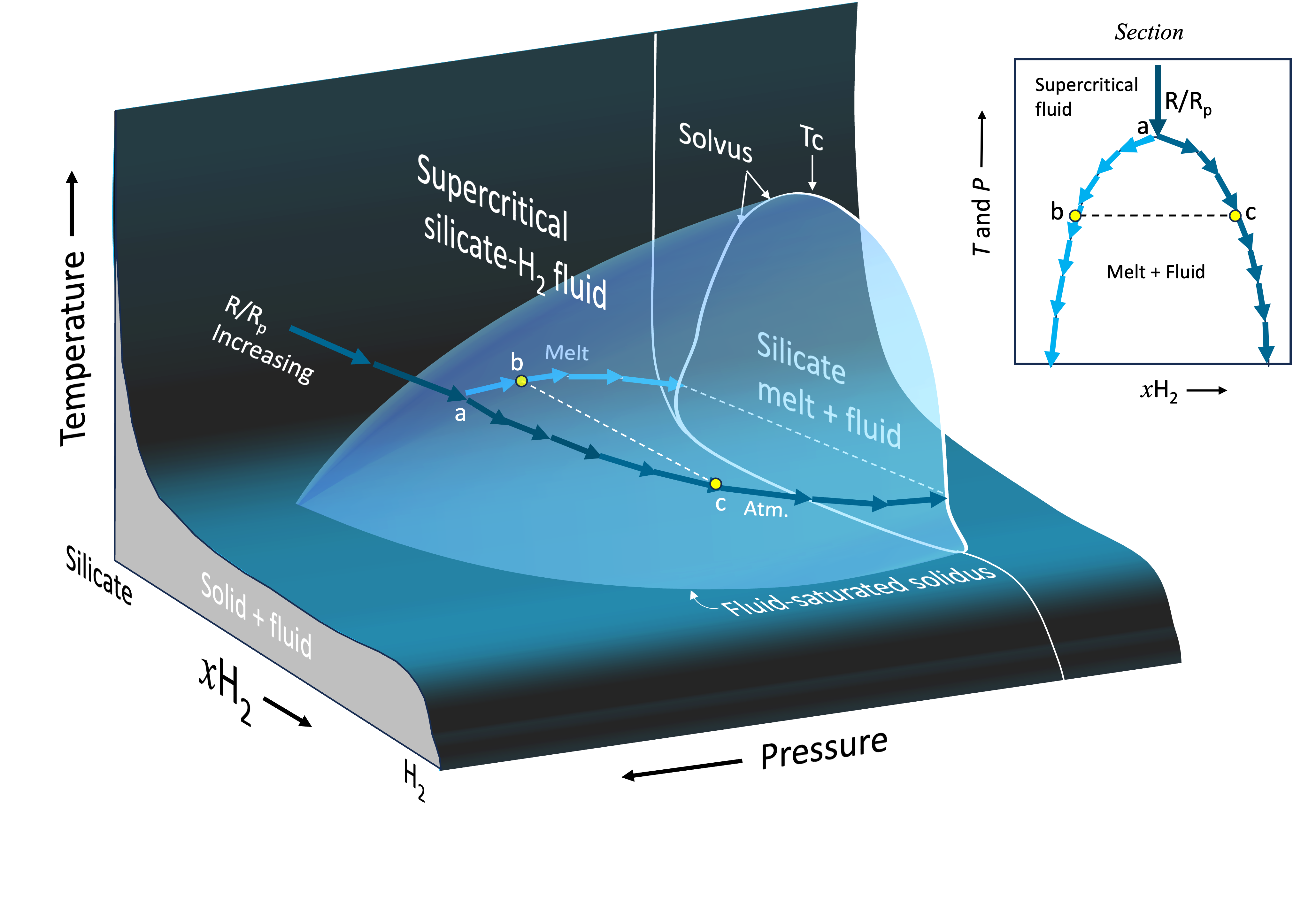}
    \caption{A schematic representation of a possible topology for the silicate-H$_2$ system as a function of temperature and pressure, adapted from the peridotite-water system as portrayed by \cite{Mibe2007}. Here, the composition axis, $x{\rm H}_2$, represents the mole fraction of H$_2$. Arrows show an illustrative path through the planet towards the surface. Also shown is a 2D polybaric section corresponding to the increasing $T$ and $P$ path shown by the arrows in the 3D diagram. The arrows in both diagrams show the change in compositions of the melt and coexisting H$_2$-rich fluid once the solvus is breached. Letters 'a', 'b', and 'c' indicate corresponding compositions along the solvus.  }
\label{fig:critical_point}
\end{figure}

 The situation is more complicated than described above, however, because the solvus must be pressure dependent.  One can hypothesize, for example, that the topology of the silicate-hydrogen system might resemble that of the peridotite-water system (peridotite representing ultramafic silicate), in which the temperature of the solvus decreases with pressure (Figure \ref{fig:critical_point}). In this case, pressure gradients can cause silicate-hydrogen phase separation or homogenization that in turn define the surfaces of ``cores", or the silicate ``weather" in hydrogen-rich envelopes. 

 The distribution of volatile species between primary atmospheres and magma oceans is another critical issue that affects the chemistry and structure of rocky planets.  For example, it is clear that water is not only partitioned according to solubilities \citep[e.g.,][]{Dorn2021}, but is also manufactured by reactions between silicates and H$_2$ \citep{Ikoma2006, Kite_2020b, Kite2021, Schlichting_Young_2022, Young2023a}. 

\section{Implications of silicate-H$_2$ miscibility}
\label{section:implications}

\subsection{Atmosphere structure}
The existence of the a silicate-hydrogen solvus not only changes the definition of the AMOI, but also the structure and chemistry of the atmosphere.  Regardless of the total entropy of the core, the temperature and pressure at the AMOI will be defined by the solvus.  One can think of the surface of the supercritical magma ocean and its interface with the overlying hydrogen-rich envelope as an isotherm for a given pressure, the location of which depends on the thermal state of the planet at a specified time.  It is the radial {\it position} of the AMOI surface that will depend on mass and time, rather than the $T$ and $P$ at the AMOI. 

The structure of the atmosphere is influenced by silicate-hydrogen miscibility because saturation at the surface of the supercritical magma ocean is maintained upward in the atmosphere as the silicate-rich and hydrogen-rich phases continue to unmix from one another with decreasing  $T$ and $P$ (e.g., Figure \ref{fig:critical_point}). This would not be the case for a simple phase change that is univariant in $T$ versus $P$ space. 
 To see this, consider application of Gibbs phase rule to our binary system: $f=c-p+2$ where $c$ is the number of components defining the chemical system, $p$ is the number of coexisting phases, and $f$ is the degrees of freedom among intensive variables.  For two phases in a binary chemical system, the degrees of freedom for intensive variables is 2, meaning that the two exsolving phases coexist over a range of temperatures and pressures. This is evident from inspection of Figure \ref{fig:critical_point}. Most paths starting in the supercritical region of the phase diagram in which $T$ and $P$ both decrease will lead inexorably to contact with the solvus comprising a hull in Figure \ref{fig:critical_point}.  We therefore have some leeway in what follows to ascribe changes in compositions of the two phases in our isobaric phase diagram to the effects of both $T$ and $P$ in the atmospheres. We note, however, that as pressure decreases upward in the atmosphere, the temperatures corresponding to the solvus may increase (e.g., Figure \ref{fig:critical_point}).  This would have the effect of driving the condensing melt towards MgSiO$_3$ and the host vapor towards H$_2$.  

A self-consistent treatment of planet formation and evolution that includes shifts in the radial position of the solvus with time is postponed for future work.  For now, for simplicity of presentation, we will continue to represent the position of the solvus as being at $R_{\rm p} = R_c \propto M_c^{\beta}$ with $\beta=1/4$ \citep{valencia2006}, mindful that $R_{\rm p}$ actually varies, by definition, with the position of the solvus.

We can illustrate the structure of the atmosphere above first contact of the $T$-$P$ path with the solvus by using the phase diagram in Figure \ref{fig:solvus1percent} (appearing in Figure \ref{fig:solvus10percent} as well) to define the composition of the gas and the melt condensing from the gas, with the lower boundary conditions for the atmosphere temperature and pressure set by the first contact with the solvus (e.g., point $a$ in Figure \ref{fig:critical_point}).  Analogous to the case for the core in \S \ref{section:time_evolution}, the structure of the atmosphere is obtained by integrating the equations,

\begin{equation}
    \frac{dm_{\rm Atm}}{dr}=4 \pi r^2 \rho,
\label{eq:dmdr_atm}
\end{equation}
\begin{equation}
    \frac{dP}{dr}=-\frac{Gm \rho}{r^2},
\label{eq:dPdr_atm}
\end{equation}
and,
\begin{equation}
    \frac{dT}{dr}=\frac{\nabla_{\mathrm{rad}} \nabla_{\mathrm{conv}}}{\nabla_{\mathrm{rad}}+\nabla_{\mathrm{conv}}}. 
    \label{eq_effectivemass_atm}
\end{equation}
\vspace{6pt}

\noindent where $m_{\rm Atm}$ is the mass of the atmosphere, $m$ is the total mass contained within radius $r$, and the numerical integration is from the magma ocean outward. The ideal gas law is used for the equation of state relating density and pressure (Equation \ref{eq:idealgas}).

\begin{figure*}
\centering
   \includegraphics[width=0.7\textwidth]{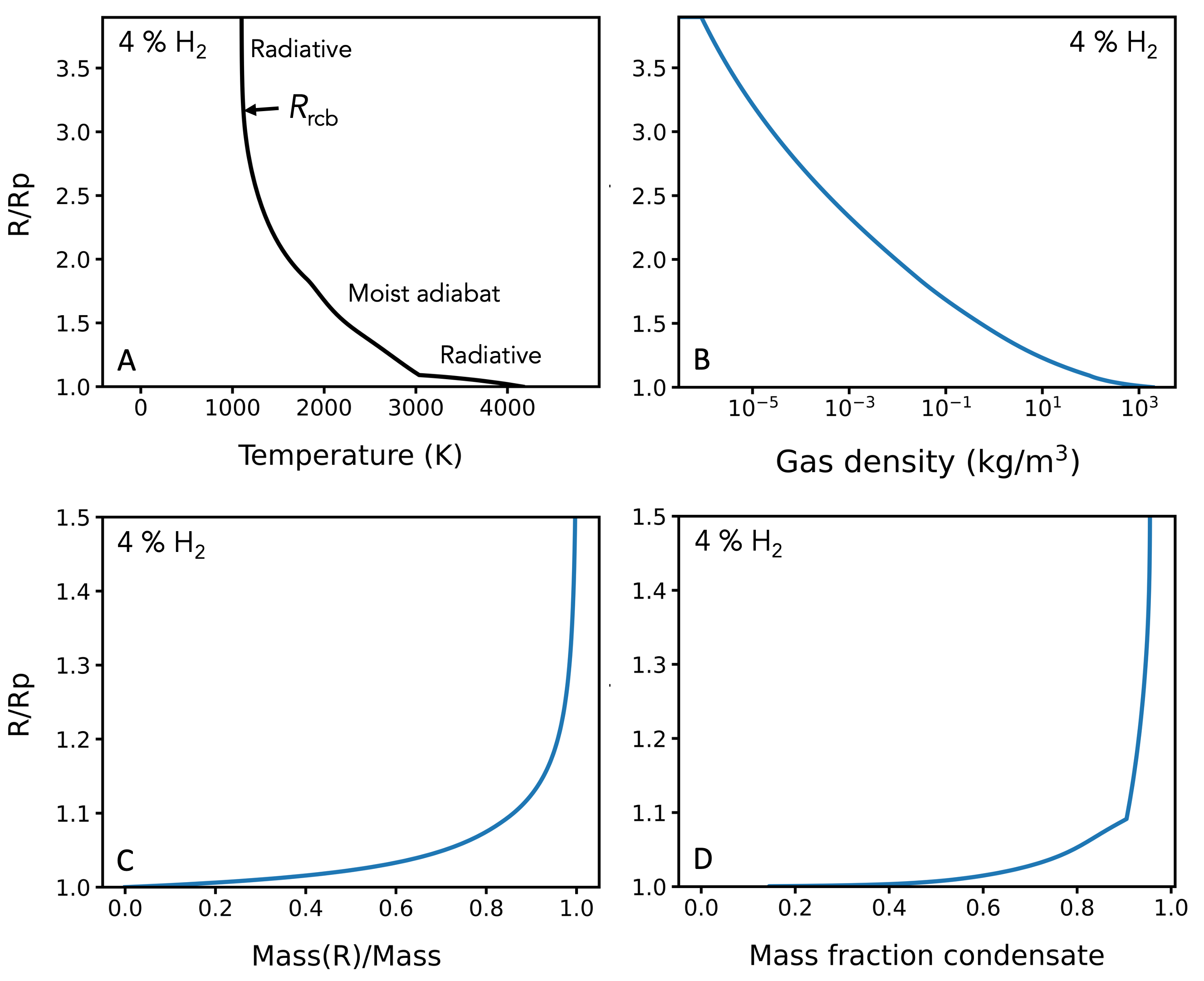}
    \caption{  Atmosphere  for the case of a $4M_{\oplus}$ planet where the bulk composition of the system is 4\% by mass H$_2$.  The calculations is based on an equilibrium temperature of $400$ K, resulting in an effective radiation temperature of 1,400 K. A) Temperature vs normalized radial position above the magma ocean where $R$ is radial distance from the center and $R_{\rm p}$ is the core radius, showing the three distinct regions: a radiative layer adjacent the surface of the magma ocean where $dT/dr$ is superadiabatic;  a convective region corresponding to a moist adiabat; and the radiative layer above the radiative-convective boundary, $R_{\rm rcb}$,  at $R/R_{\rm p} = 3.1$.  B) Density profile for the vapor phase.  C) Fraction of atmosphere mass contained within radial positon $R/R_{\rm p}$.  D) Mass fraction of the atmosphere composed of silicate condensate.  The gaseous atmosphere comprises 2.24\% of the planet by mass, with the remainder of the hydrogen in the condensates and in the interior of the planet.   }
\label{fig:T_vs_R_4percent}
\end{figure*}

Because the vapor and silicate melt are in equilibrium, we use the moist pseudoadiabat of \cite{Graham_2021} to evaluate $\nabla_{\mathrm{conv}}$ in Equation \ref{eq_effectivemass_atm}. Molar heat capacities required to evaluate the pseudoadiabat from \cite{Graham_2021} are obtained from the NIST thermodynamic data base \citep{Chase1998} where we assume the vapor silicate component speciates to SiO, Mg, and O$_2$. Hindrance of convection due to the mass load of heavy elements at relatively high temperatures must be included, as described previously for similar circumstances by \cite{Misener2023}.  

By including the Ledoux criterion rather than just the Schwarzschild criterion for convection, the calculations permit development of a radiative layer at the base of the atmosphere that transports the intrinsic heat coming from the magma ocean upward \citep{Guillot2010,Misener2023}.  The intrinsic luminosity due to the heat emanating from the core to space becomes a determining factor for the structure of the atmosphere. Accordingly, we use Equation \ref{eq_effectivemass} as before to calculate temperature gradients, but with modifications. Since Eddington's equation for radiative diffusion relies on the local luminosity, the temperature gradient due to radiative diffusion includes the radially-dependent luminosity, $L(r)$.  At depths well below the radiative-convective boundary, stellar radiation cannot penetrate the atmosphere.  Therefore, at these depths the radiative diffusion is given by

\begin{equation}
    \nabla_{\rm rad}=-\frac{3 \kappa \rho L(r)}{64 \pi \sigma T^3 r^2},
    \label{eq_nablarad_r}
\end{equation}
where $L(r)$ is related to the intrinsic luminosity, $L_{\rm int}$, by $L(r)=(L_{\rm int}/\tau) (r^2/R_{\rm B}^2)$ and $\tau$ is the optical depth evaluated in our numerical calculation as $\kappa(r)\, \rho(r)\, dr$ for each layer of thickness $dr$.  The optical depth accounts for the mean free path of photons.  The intrinsic luminosity is manifested by the outgoing thermal radiation at the top of the atmosphere as the difference between the effective radiation temperature and the equilibrium temperature,  $L_{\rm int}=\sigma (T^4_{\rm eff}-T^4_{\rm eq})4\pi R_{\rm B}^2$.

We use the criterion for convective inhibition given by \cite{Markham2022}, expressed  in terms of the mole fraction of heavy elements relative to H$_2$, $1-x{\rm H}_2$:

\begin{equation}
(1-x{\rm H}_2)_{\rm critical}=1/\left((\Delta \hat{H}_{\rm c}/(RT)-1)(\epsilon-1)\right)
\label{eq_Markham_criterion}
\end{equation}
where $\Delta \hat{H}_{\rm c}$ is the latent heat of condensation per mole of condensate and $\epsilon$ is the ratio of the mean molecular weight of the condensable gas  to H$_2$ where we assume MgSiO$_3$ in the gas phase instantly speciates to SiO, Mg and O$_2$. The heat of condensation is calculated as $\Delta \hat{H}_{\rm c}=$ $\hat{H}_{\rm MgSiO_3} - \hat{H}_{\rm SiO} - \hat{H}_{\rm Mg} -\hat{H}_{\rm O_2}$ where the poorly known effect of H dissolved in the condensate on the latent heat is assumed to be of minor importance.  Molar enthalpies as a function of temperature are obtained from the NIST thermodynamic data base \citep{Chase1998}.  Where the mole fraction of heavy elements exceeds $(1-x{\rm H}_2)_{\rm critical}$ the temperature gradient is given by Equation \ref{eq_nablarad_r}.

Results for the case of a $4M_{\oplus}$ planet  where the bulk composition of the system, at least near the surface, is 4\% by mass H$_2$, is shown in Figure \ref{fig:T_vs_R_4percent}.  The temperature profile exhibits a radiative layer adjacent the surface of the magma ocean where $\nabla T$ is markedly superadiabatic due to the heavy load of silicates in the saturated vapor phase.  Above this layer is a convective region corresponding to a moist adiabat, and above that is the radiative-convective boundary at $R/R_{\rm p} = 3.1$.  These calculations were performed assuming the pressure for the AMOI (i.e., the solvus) is 5 GPa.  The pressure effects on solvus temperatures are beyond the scope of this present work but are roughly constrained to be relatively modest at GPa pressures by the simulations by \cite{Gilmore2023}.  The mass of the atmosphere in this example is $2.24$\% of the planet, with most of the hydrogen residing in the condensed core. Most of the mass of the atmosphere resides in the lowest radiative layer where convection is inhibited by the gradient in molecular weight (Figure \ref{fig:T_vs_R_4percent}).  

The atmosphere in this example is unlike those depicted in Figures \ref{fig_structure} and \ref{fig_coresurface}. For example, consider the position of a $\sim 2$\% by mass hydrogen atmosphere with a basal temperature about 4,000 K and a basal pressure of 5 GPa in Figures \ref{fig_structure} and \ref{fig_coresurface}.  The position corresponding to these conditions lies at the lowest part of the evolution curves in the figures, implying  that the $4M_{\oplus}$ planet had experienced extensive cooling and mass loss. However, the conditions and masses of the atmospheres in the evolution paths depicted in Figures \ref{fig_structure} and \ref{fig_coresurface} are replaced by exsolution of the silicate-rich and H$_2$-rich phases along their solvus.  The $T$, $P$, and mass of the atmosphere are therefore not directly reflective of the cooling and/or mass-loss path taken by the planet.  Rather, the radial {\it position} of the solvus would be affected by the accretion, cooling, and mass loss history. 

The radius of the planet is affected by the solvus.  In the present example,  assuming a silicate mantle and iron core in Earth-like proportions, the radius defined by the line-of-sight optical depth (chord optical depth) of unity for H$_2$ opacities is $3.5 R_{\oplus}$. For comparison, using the methods described in  \S \ref{section:TandP} for a planet with the same mass of H$_2$ atmosphere ($2.2$\%), a similar $T_{\rm eff}$ of 1,010 K, and a comparable surface temperature of 3,900 K, $R_{\rm rcb}/R_{\rm p}$ in the latter case is  $\sim 2$ compared with $\sim 3$ in the model in Figure \ref{fig:T_vs_R_4percent}. The transit radius is therefore smaller, with a value of $2.2 R_{\oplus}$. Differences in radii due to density deficits in the silicate and metal cores are important, but relatively small in comparison to the effects of the solvus on the structure of the atmosphere \citep{Rogers2024}. Evolutionary calculations will be needed to produce more meaningful comparisons between models with and without the silicate-hydrogen solvus controlling conditions at AMOIs.    

\subsection{Detectability: silicate rain or fog?}

The structure and composition of the atmosphere, and thus the ability to detect the chemical signatures of a magma ocean, depends on whether condensing silicate rains out of the atmosphere or remains suspended, that is, does it form rain or fog? Efficacious rainout at all depths in the atmosphere would drive the vapor phase to a pure H$_2$ composition by distillation, scrubbing it of silicate, and returning the silicate to the magma ocean below.  This in turn would render detection of silicate in the atmosphere difficult if not impossible.  It would also limit convective inhibition by decreasing the silicate load in the atmosphere, thus altering the transit radius of the planet.  

Evaporation of condensed droplets will be matched by condensation at steady state by virtue of the effectively infinite supply of silicate from the magma ocean.  Since  silicate will continue to exsolve from the gas phase (Figure \ref{fig:solvus10percent}) we do not consider the longevity of droplets in the aggregate to be controlled by evaporation \citep[c.f.,][]{Loftus2021}. Rather, we address the fate of silicate condensates by comparing the absolute values of their terminal velocities, $v_{\rm t}$, to net vertical turbulent velocities, $u$, in atmospheres. Where terminal velocities exceed turbulent velocities, so that $v_{\rm t}/u > 1$, silicate condensate droplets can fall as precipitation.  Where turbulent velocities exceed terminal velocities, with $v_{\rm t}/u < 1$, condensates can be suspended in the atmosphere. We use the term ``fog" in the latter case because the existence of the silicate-hydrogen solvus dictates that the vapor phases is saturated in silicate at ``ground level" at the surface of the magma ocean. Saturation persists upward with decreasing temperatures as long as the supply of silicate persists (Figure \ref{fig:solvus10percent}). As shown below, terminal velocities increase with height in the atmosphere while turbulent velocities decrease (for a fixed vertical eddy diffusivity), so suspension of droplets is favored at ground level near the solvus; condensates suspended in a turbulent atmosphere are favored near ground level where the magma ocean provides a steady source of silicate.  

Estimates for the sizes of silicate condensate droplets are required to determine their fate.  Maximum radii for droplets occur because drag forces from the enveloping gas eventually overwhelm droplet surface tension.  We estimate these maximum droplet radii using the Weber number, $\mathit{We}$, the  dimensionless ratio of inertial forces (drag) to surface tension, $\sigma_c$, in combination with estimates for the headwind velocity, $v$, encountered by the droplets. In this approach, the maximum radii for condensate droplets against disruption by drag forces is given by 

\begin{equation}
r_{\rm max}=\frac{{\mathit{We}}\, \sigma_c}{\rho v^2},
\label{eq:rmax}
\end{equation}
where $\rho$ is the density of the gas medium and the critical value for $\mathit{We}$ us usually taken to be 4 \citep{Loftus2021}. We  evaluate Equation \ref{eq:rmax} with $v = v_{\rm t} + u$, so that the maximum radii of the droplets is set by disruption due to turbulence where terminal velocities are negligible, and vice versa.  The turbulent velocity in the vertical direction can be obtained from \citep[e.g.,][]{Swenson2021}

\begin{equation}
u(r) = \frac{K_{zz}}{H(r)},
\label{eq:vz}
\end{equation}
where $H(r) = k_{\rm B}T/(\mu(r) g(r))$ is the scale height for the atmosphere with mean molecular mass $\mu(r)$ and $K_{zz}$ is the vertical eddy diffusion coefficient (m$^2$/s). Estimates for $K_{zz}$ based on observations of sub-Neptunes \citep{Blain2021} and theory \citep[e.g.,][]{Youdin_2010} suggest appropriate values for $K_{zz}$ are likely between 10 and $10^5$ m$^2$/s.  

Condensate droplets exsolving from the supercritical fluid will experience viscous drag that controls their terminal velocities through the gas.  The drag force on a spherical droplet, $F_{\rm D}$, in this case is  

\begin{equation}
F_{\rm D}=4\pi r_{\rm d}^2 \frac{C_{\rm D}}{2} {\rho} v_{\rm t}^2,
\label{eqn:dragforce}
\end{equation}
where ${\rho}$ is the density of the surrounding gas, $r_{\rm d}$ is the radius of the droplet sphere, here taken to be $r_{\rm max}$, $v_{\rm t}$ is the velocity relative to the enveloping gas, and $C_{\rm D}$ is the drag coefficient.  The drag coefficient for turbulent flow with a no-slip condition at the surface of the drops is \citep{Feng2001}

\begin{equation}
C_{\rm D} =\frac{24}{\rm Re}\left(1+{\frac{1}{6}}{\rm Re}^{2/3}\right),
\label{eq:dragcoefficient}
\end{equation}
where Re is the Reynolds number such that ${\rm Re} = 2 v_{\rm t} r_{\rm d}/\nu$ and $\nu$ is the kinematic viscosity for the gas. 

The drag force is combined with the buoyancy force, $F_{\rm B} = (4/3)\pi r^3 \overline{\rho} g$,  and the force of gravity, $F_{\rm g}= (4/3)\pi r^3 \rho_{\rm d} g$, to yield the force balance $F_{\rm D} + F_{\rm B} = F_{\rm g}$.  This balance can be solved for the velocity of the condensates:

\begin{equation}
v_{\rm t} = \sqrt{\frac{2}{3}\frac{r_{\rm d} (\rho_{\rm d}-{\rho})g}{C_{\rm D}{\rho}}.
}
\label{eq:velocity}
\end{equation}

\begin{figure}
\centering
   \includegraphics[width=0.41\textwidth]{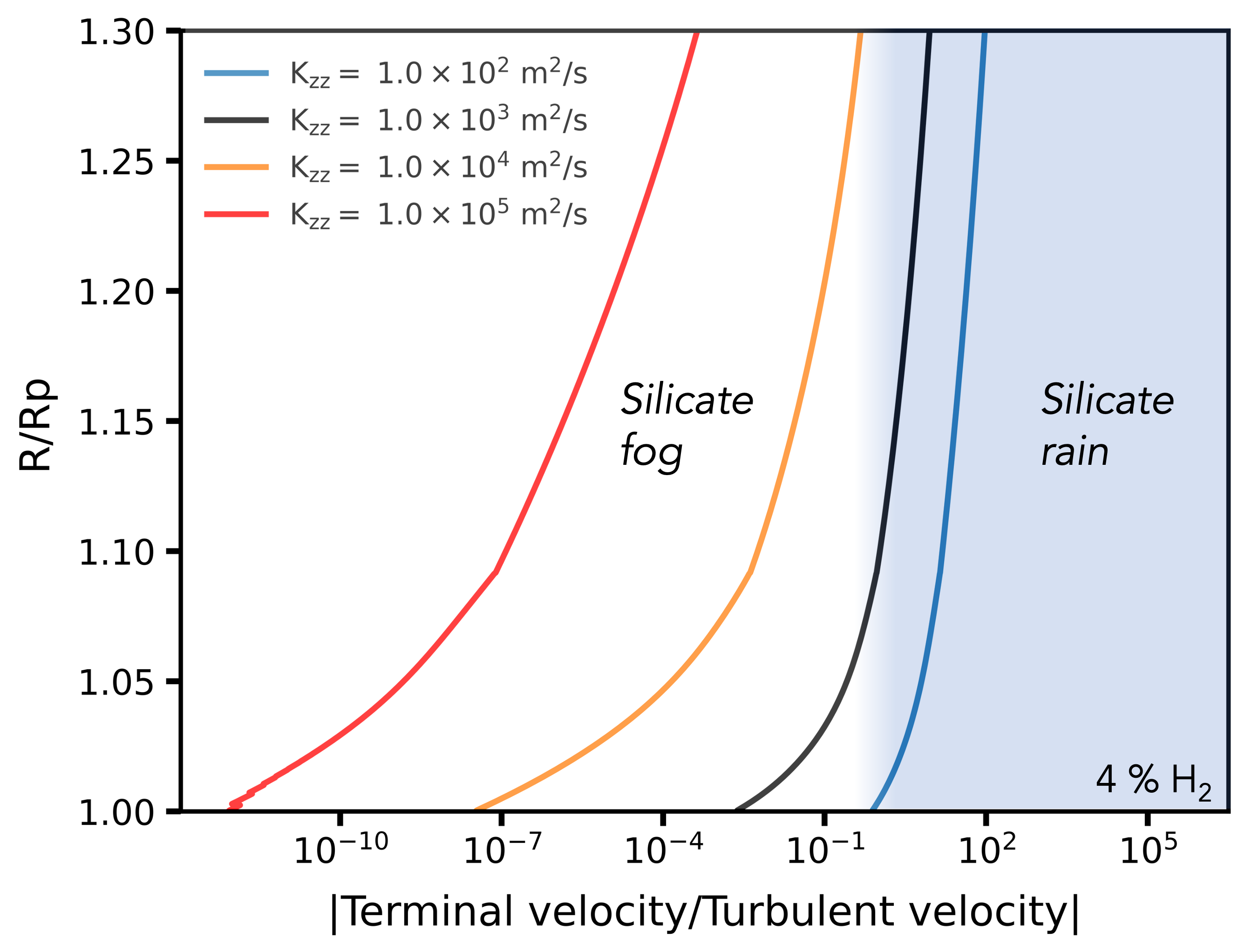}
    \caption{  Terminal velocities relative to turbulent velocities of silicate condensates in the atmosphere shown in Figure \ref{fig:T_vs_R_4percent} based on Equations \ref{eq:rmax}, \ref{eq:vz}, \ref{eq:dragcoefficient}, and \ref{eq:velocity}. Results for four different values for Kzz are shown. Shading demarcates the boundary between falling rain and ``fog" where condensates are entrained by the vertical component of turbulence. Velocity ratios are shown as absolute values, ignoring the opposing directions.   }
\label{fig:rain_vs_fog_velocities}
\end{figure}

We solved Equations \ref{eq:rmax}, \ref{eq:vz}, \ref{eq:dragcoefficient}, and \ref{eq:velocity}  simultaneously by iteration to yield self-consistent models for turbulent drag in terms of Reynolds number for spheres of silicate melt descending through the gas phase of the atmosphere model in Figure \ref{fig:T_vs_R_4percent}. We used the equation of state for enstatite to approximate MgSiO$_3$-rich melt densities and a high-$T$, high-$P$ surface tension $\sigma_c$ of $0.5$ N/m for the melt \citep{Walker1981, Colucci2016} (results are insensitive to the range of published values).  We calculate kinematic viscosities of the atmosphere gas from dynamic viscosities of H$_2$, assuming the gas behaves like H$_2$, and gas densities at each elevation.  The dynamic viscosities of H$_2$ are obtained from the temperature-dependent Sutherland constants for hydrogen gas from \cite{Braun2018} and the densities shown in Figure \ref{fig:T_vs_R_4percent}. 

\begin{figure}
\centering
   \includegraphics[width=0.41\textwidth]{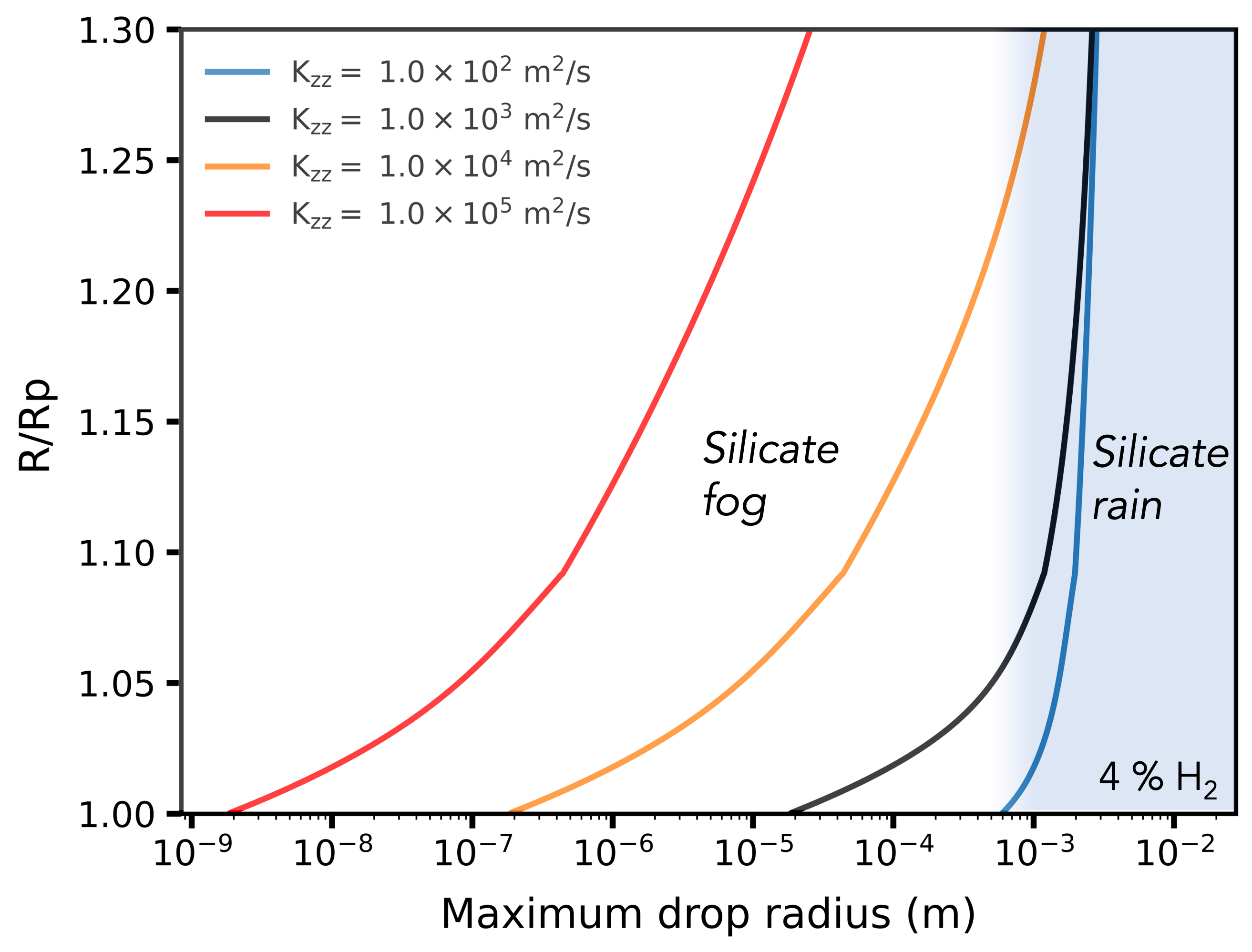}
    \caption{  Maximum droplet radii for silicate condensates in the atmosphere shown in Figure \ref{fig:T_vs_R_4percent} based on Equations \ref{eq:rmax}, \ref{eq:vz}, \ref{eq:dragcoefficient}, and \ref{eq:velocity}. Results for four different values for Kzz are shown. Shading shows the boundary between falling rain and ``fog" where condensates are entrained by the vertical component of turbulence.   }
\label{fig:rain_vs_fog_radii}
\end{figure}

Results (Figures \ref{fig:rain_vs_fog_velocities} and \ref{fig:rain_vs_fog_radii}) show that if eddy diffusivities, $K_{zz}$, are greater than $\sim 10^2$, droplets of silicate condensate may be suspended in the basal radiative layer. For example, following the contour for $K_{zz} = 10^3$ downward in $R/R_{\rm p}$ in Figure \ref{fig:rain_vs_fog_radii}, maximum drop radii near $10^{-3}$ m decrease to $\sim 5\times 10^{-4}$ m at $R/R_{\rm p} \sim 1.06$ due to disruption as they travel through the gas phase.  Below this level, these smaller droplets are suspended in the gas phase (Figure \ref{fig:rain_vs_fog_velocities}). For this degree of turbulence, rain drops are dispersed as they descend to become ``fog".  

For $K_{zz} \ge 10^4$, condensates have the opportunity to be suspended in the atmosphere well into the convective layer (Figures \ref{fig:rain_vs_fog_velocities} and \ref{fig:rain_vs_fog_radii}). At still greater values for $K_{zz}$, condensates are predicted to be fully entrained in the atmosphere with no rainout. Under these circumstances, the vapor phase may remain saturated and silicates derived from the magma ocean may be detectable high in the atmosphere.  In reality, trajectories of droplets in Figures \ref{fig:rain_vs_fog_velocities} and \ref{fig:rain_vs_fog_radii} will cut across $K_{zz}$ contours as the drivers of turbulence vary at different levels in the atmospheres (e.g., when crossing from radiative to convective layers).

\subsection{Molecular weight of the atmospheres}
The atmospheres produced by existence of a silicate-hydrogen solvus will have relatively high molecular weights at high temperatures near the base, and progressively lower molecular weights higher up where temperatures decrease.  In the example shown in Figure \ref{fig:T_vs_R_4percent}, the molecular weight of the vapor phase is $13.3$ g/mole just above the surface of the magma ocean, and the molecular weight of the condensed phase is $39.5$ g/mole; the two phases differ in molecular weight by just a factor of 3 rather than the factor of 50 for pure MgSiO$_3$ and H$_2$.  Outward from the surface the molecular weight of the vapor phases decreases and the molecular weight of the condensate increases, as prescribed by the shape of the solvus. The result is that above the convectively inhibited radiative layer, at $R/R_{\rm p} > 1.06$, silicate condensates dominate the mass fraction of the atmosphere. 

\begin{figure*}
\centering
   \includegraphics[width=0.85\textwidth]{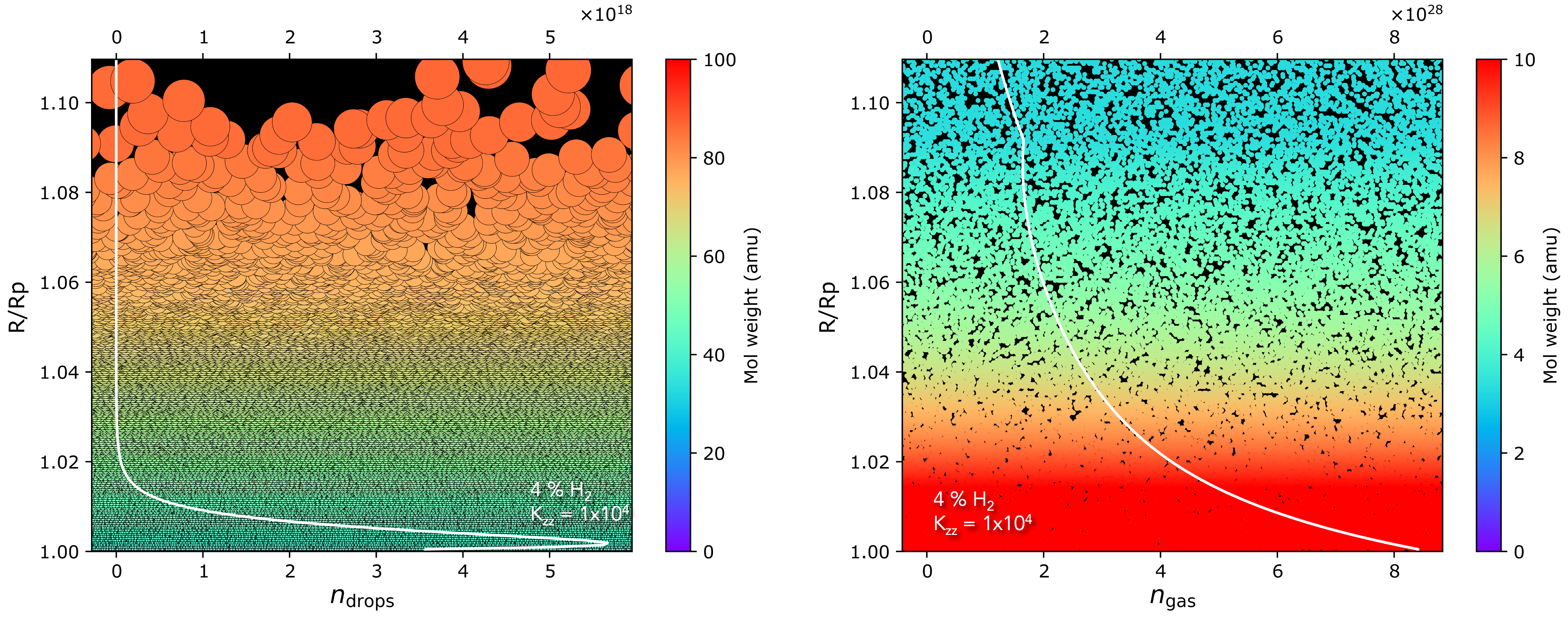}
    \caption{  Number densities of condensate droplets and gas molecules, and their respective molecular weights, as a function of height $R/R_{\rm p}$ for the atmosphere shown in Figure \ref{fig:T_vs_R_4percent} with $K_{zz} = 1\times 10^{4}$. Left: Numbers of dots at each height in the atmosphere represent the relative number densities of droplets, and the symbol sizes are proportional to droplet sizes. Right: Numbers of dots represent the relative number densities of molecules. In both plots, colors indicate molecular weights at each height (note different color scales in the two panels).   Curves for absolute number densities of droplets (left) and gas molecules (right) are overlain for comparison.    }
\label{fig:rendering}
\end{figure*}

It is arguably difficult to visualize mass fractions of species with such large differences in molecular weight, so we recast the concentrations of condensate into number densities and volume fractions. The gradients in molecular weight are shown in Figure \ref{fig:rendering} for the case were $K_{zz} = 1\times 10^4$.  Here, in the case of droplets, the number of dots at each height in the atmosphere are in proportion to the number densities of droplets and their size is proportional to the size of the droplets. In the case of the gas phase, the number of dots is proportional to the number density for the molecules.  In both cases, each dot is color-coded for the molecular weight (melt) or mean molecular weight (gas), where in the case of the gas, speciation to SiO + Mg + O$_2$ is assumed. Curves showing absolute number densities as a function of height $R/R_{\rm p}$ are overlain for comparison.  Number densities for the condensate droplets are obtained from the maximum radii (Figure \ref{fig:rain_vs_fog_radii}). The number density of condensate droplets is 14 orders of magnitude smaller than that of the molecules comprising the gas near the top of the convectively inhibited layer, above which there is a sharp drop off in number density into the convective layer (Figure \ref{fig:T_vs_R_4percent}) even as the molecular weight of the condensates approaches that of pure MgSiO$_3$.  The number density and size of droplets correspond to volume-filling fractions for the atmosphere (analogous to liquid water content, or LWC in the terrestrial atmosphere) of about 20\% at this level.  This is indeed a dense fog, as it is roughly five orders of magnitude greater than that for a thick water fog in the terrestrial atmosphere.   

The gas phase exhibits a sharp gradation in molecular weight, with values 5 times that of pure hydrogen just above the magma ocean.  Where the lower radiative layer gives way to convection at $R/R_{\rm p} \sim 1.06$, the molecular weight of the vapor phase is still well above that of pure H$_2$.  However, at the height of the $R_{\rm rcb}$ (above the maximum in Figure \ref{fig:rendering}), the molecular weight of the gas is essentially that of hydrogen in this model.  Silicate exists at trace levels, which may or may not be detectable depending on the circumstances. The present simple model is inadequate to capture the critical details of trace amounts of condensables.

It is straightforward to layer onto the calculations above the expected equilibrium concenstrations of H$_2$O and SiH$_4$ using equilibrium constants among the gas species, as in \cite{Misener2023}.  Including these species does not change the salient features of the model described above.

 \subsection{ The compositions and densities of metal cores}

The high concentrations of hydrogen in silicate melt implied by the miscibility of silicate and hydrogen has consequences for the formation of metal cores.  High-temperature reactions between cores and hydrogen-rich atmospheres are expected to greatly affect the compositions of metal cores senso stricto.  \cite{Schlichting_Young_2022} used extrapolations of existing thermochemical data to suggest that at relevant temperatures, metal cores should acquire significant fractions of H as well as Si, O, and, by inference, perhaps some other light elements.  It has been proposed that the density of Earth's core was determined by this process \citep{Young2023a}, albeit at less extreme conditions. These first results raise the possibility that not all rocky planets form discrete metal cores.  If chemical equilibrium is rapid at the atmosphere magma-ocean interface, substantial masses of H enter Fe-rich metals, along with Si, O, and other light elements.  The precise compositions depend critically on equilibration temperatures, but at T $\ge$ 3,000 K, our existing chemical equilibrium models \citep[e.g.,][]{Schlichting_Young_2022, Young2023a}  suggest that light elements can introduce density deficits approaching 50\% or greater.  While the stoichiometries of these alloys resemble oxides and hydrides (e.g., Fe$_{0.2}$ Si$_{0.1}$ H$_{0.6}$ O$_{0.1}$), previous ab initio calculations suggest that they may retain a metallic character \citep{Scipioni2017}, comprising a phase distinct from silicates and oxides even in a molten state.  

Even where a discrete metal phase is assured, light elements in metals affect the prospects for forming a discrete metal core at the center of a planet.  A self-consistent model for turbulent drag for liquid metal within the silicate magma oceans suggests that differentiation into a discrete metal core may not be feasible. 

We assume that the silicate magma oceans are generally turbulent due to the high Rayleigh numbers, ${\rm Ra}$, where ${\rm Ra} = (g \alpha \Delta T L^3)/(\kappa \nu)$, and where here $\kappa$ is the thermal diffusivity for the melt, $\nu$ is the kinematic viscosity, $L$ is the depth of the magma ocean, $\Delta T$ is the change in temperature from the base to the surface of the magma ocean, and $\alpha$ is the expansivity.  Using typical values for silicate melt, ${\rm Ra} \sim 10^{30}$ for our fiducial 4$M_{\oplus}$ planet, which is more than sufficient for vigorous convection by many orders of magnitude. Metal segregating from the silicate will therefore not experience Stokes settling, which would imply a laminar enveloping flow field, but rather viscous drag.   
The drag force on a spherical segregation of metal, $F_{\rm D}$, and the associated velocity, is in this case given by Equations \ref{eqn:dragforce}, \ref{eq:dragcoefficient}, and \ref{eq:velocity} where now the medium is silicate and the spherical body passing through the medium is metal. In this application, ${\rho}$ is the density of the surrounding medium, $\rho_{\rm d}$ is the density of the molten metal, $r_{\rm d}$ is the radius of the molten metal sphere, and $v_{\rm t}$ is the velocity of the metal sphere relative to the silicate melt.   

\begin{figure*}
\centering
   \includegraphics[width=0.85\textwidth]{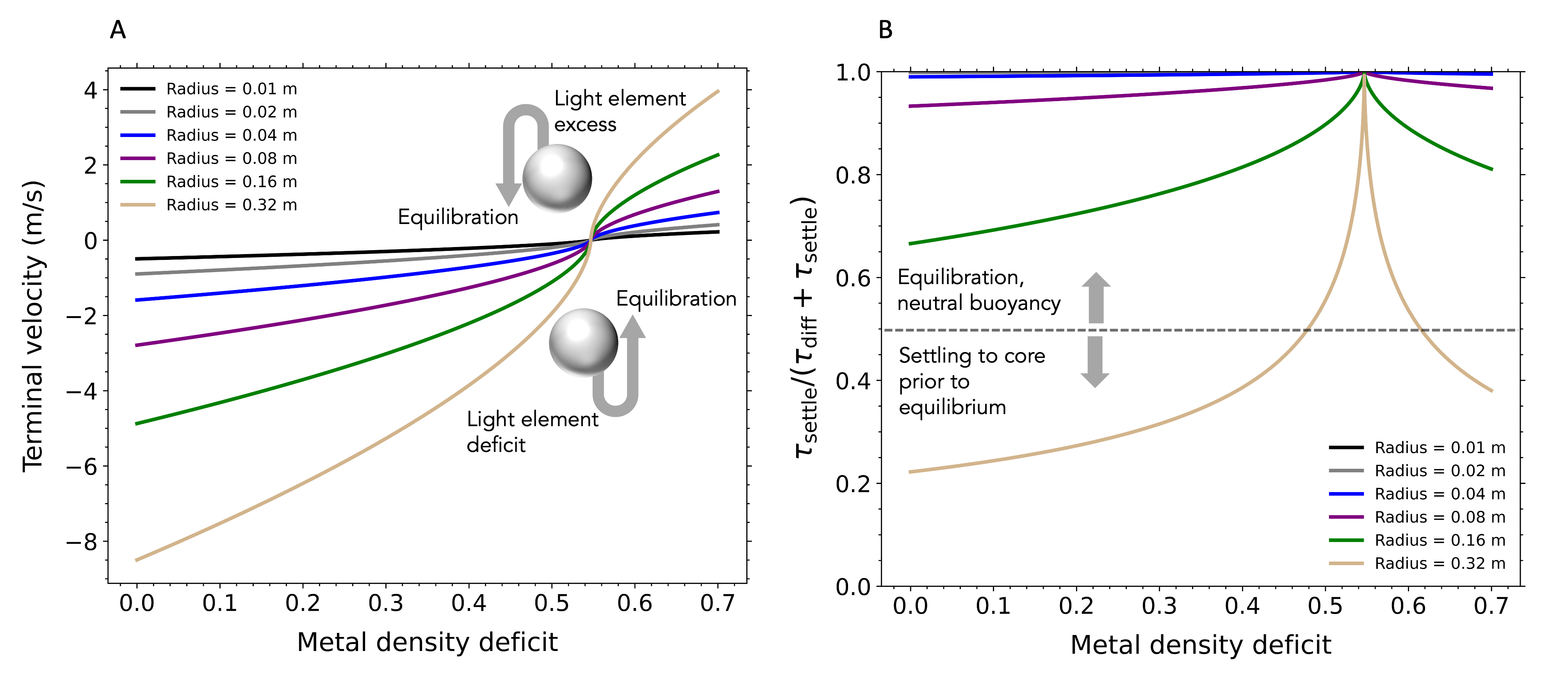}
    \caption{A: Terminal velocities for Fe metal droplets of different radii settling through silicate melt as a function of density deficits due to light elements in the metal.  The light element concentrations in the metal depend on temperature \citep{Schlichting_Young_2022}, and thus depth. The curves correspond to a fixed temperature of 5,000 K, and a pressure of 10 GPa.  For a given density deficit, the calculation is not critically dependent on the precise $T$ and $P$.  For comparison, melt convective velocities should be $<$1 m/s \citep{Solomatov2009,Young2023a}. B: Settling versus diffusion timescales as a function of metal density deficit and radius. For values of $\tau_{\rm settle}/(\tau_{\rm diff} + \tau_{\rm settle})< 0.5$, metal droplets can settle before achieving equilibrium concentrations of H.  For values  $> 0.5$, equilibration occurs prior to settling to the core. }
\label{fig:terminal}
\end{figure*}

Here again, 
because $C_{\rm D}$ depends on Re, Re depends on $v_{\rm t}$, and $v_{t}$ on $C_{\rm D}$, self consistent solutions are obtained for viscosity, Reynolds number, and velocity by iteration. This self-consistent model for turbulent drag in terms of Reynolds number for a sphere of liquid metal alloy descending through a molten silicate magma ocean yield, for example, Re = $63100$, C$_{D} = 0.101$, and $\nu = 2.9\times 10^{-5}$ ${\rm m}^2$/s, corresponding to a dynamic viscosity $\eta$ of $0.1$ Pa s, for an alloy sphere 50 cm in radius with a 50\% density deficit. Solutions for a variety of radii and denstiy deficits (Figure \ref{fig:terminal}A) show that a point of neutral buoyancy is obtained at density deficits of about $55\%$ using the nominal equations of state for silicate and alloy in the Appendix. 

Analysis of metal droplet sizes stable against disruption suggest that larger masses of metal settling through silicate melt will break up until they reach diameters of $\sim 10^{-2}$ meters \citep{Rubie2003}.  Metal spheres $<$ cm-sized will be well coupled with the turbulent magma ocean, frustrating metal core formation \citep{Lichtenberg2021b}.  Here we find that cm and larger metal droplets will settle to a radial level in the body determined by the amount of alloying light elements (especially H) (Figure \ref{fig:terminal}A).  Rising above the neutral buoyancy level results in lower temperatures of equilibration, fewer light elements, and an increase in density.  Sinking below the point of neutral buoyancy results in higher temperatures of equilibration, more light elements in the metal, and lower density. Under conditions of high-$T$ equilibration between magma oceans and H$_2$-rich atmospheres, therefore, metals may become stranded at horizons corresponding to neutral buoyancy that are nearer to the surface of the bodies, hindering the formation of a central metal core. Rocky planets without metal cores were  previously postulated to exist by \cite{Elkins-Tanton_2008}, but by a different mechanism.  These authors hypothesized complete oxidation of iron metal by water to convert Fe metal to an FeO component in the silicate melt rather than incorporation of light elements into the metal phase.

Several caveats to the calculations above require evaluation. One is if the rate of exchange of light elements between the metal and surrounding silicate medium is too slow to achieve equilibrium at each depth.  One expects metal droplets to experience internal circulation due to their relatively low viscosity \citep{Dahl2010}, and thus to be well mixed.  Nonetheless, in the less likely circumstance where internal circulation is absent in metal droplets, chemical equilibration will be limited by diffusion of light elements in and out of the metal. The possibility that diffusion limits equilibration prior to settling to the core can be evaluated  by comparing the timescale for diffusive equilibration, $\tau_{\rm diff}=r_{\rm d}^2/D$, to the settling timescale $\tau_{\rm settle}=0.5R_{\rm c}/|v_{\rm t}|$, where $D$ (m$^2$/s)  is the diffusivity of a light element in metal.  We evaluate the timescale for settling using  half the radius of the planet core, $R_{\rm c}$, to accommodate buildup of an Earth-like metal core fraction.  Where $\tau_{\rm diff}$ is greater than $\tau_{\rm settle}$, metal droplets can reach a growing metal core prior to acquiring equilibrium concentrations of light elements, thus avoiding being trapped by neutral buoyancy.  

Hydrogen is by far the most effective light element for reducing the density of metal, so we evaluate $\tau_{\rm diff}$ for hydrogen. Diffusivities for H in Fe metal at the temperatures of interest here are similar at high pressures \citep{Yuan2023} and low pressures \citep{Depuydt1972} to within one order of magnitude, with values ranging from $10^{-8}$ to $10^{-7}$ m$^2$/s from 4,000 K to 10,000 K, respectively. Values for $\tau_{\rm diff}$ can be evaluated using $\log{D_{\rm H}}=-6.2 -44,000/(RT)$ in SI units, from \cite{Yuan2023}. 
For convenience the parameter $\tau_{\rm settle}/(\tau_{\rm diff} + \tau_{\rm settle})$ is used to compare timescales, ranging from 0 to 1 (settling timescales can go to zero, but diffusion timescales do not).  Values $< 0.5$ indicate metal droplets settle to the cite of metal core formation faster than they can exchange light elements with the supercritical silicate medium.  Values $> 0.5$ indicate that diffusion is sufficiently fast to allow light element equilibration prior to reaching a metal core, thus encountering neutral buoyancy. 

Results in Figure \ref{fig:terminal}B show that at 5,000 K we can expect metal droplets less than about 100 cm to reach neutral buoyancy, while larger droplets may sink too fast and reach the center of the planet if they have density deficits of less than about $0.4$.  Higher temperatures yield higher diffusivities, lower diffusive timescales, and $\tau_{\rm settle}/(\tau_{\rm diff} + \tau_{\rm settle})$ closer to 1. Larger droplets may therefore achieve  neutral buoyancy as higher temperatures are encountered with depth in the magma ocean. 

Another caveat is that the flow regime encountered by the descending droplets of metal may be less turbulent than we assert. If we assume that Stokes settling obtains, for example, the settling velocities are much greater than those based on viscous drag. In the case were diffusion limits exchange between metal droplets and the surrounding supercritical silicate medium, metal droplets greater than a few centimeters in size settling with Stokes velocities would readily reach the center of the planet without acquiring their full complement of light elements.  However, it is not clear that the laminar flow implied by Stokes settling is relevant.  Magma ocean Rayleigh numbers of order $10^{30}$ imply Reynolds numbers of $10^{15}$ using a commonly used scaling between the two for natural convection where $\mathrm{Re}\sim \mathrm{Ra}^{1/2}$ \citep{Grossmann2002}.  Such high Reynolds numbers are orders of magnitude greater than the critical value for turbulence that is on the order of $10^2$ or less \citep{WITHAM2008}.  

We repeated the calculations shown in Figure \ref{fig:terminal} using drag coefficients suitable for a no-shear boundary at the surface of the metal, replacing the the no-slip boundary \citep[e.g.,][]{Young2022_velocimetry}. The no-shear boundary might arise for the relatively low viscosity of the metal melt. The results are similar to those in Figure \ref{fig:terminal}. 

Another mitigating circumstance may be en masse motion where the mass of metal droplets exceeds a threshold. Experiments show that where the mass of the suspended phase (metal in this case) becomes high enough in the medium (silicate in this case),  en masse settling can be triggered \citep{Sparks1993}.  However, these observations do not involve a continuously changing density of the suspended phase that is capable of neutralizing the density contrast entirely.  With no density difference, the critical mass of settling phase required for rainout becomes infinite \citep[equation 1 in][]{Sparks1993}. 

One final caveat to raise is the magnitude of the density deficit of metal relative to the density of the supercritical silicate-hydrogen phase.  We do not yet have an equation of state for the latter phase, so the density deficit in the metal required to match the density of the supercritical silicate-hydrogen phase is uncertain. We used MgSiO$_3$ densities in our calculations here.  Lowering the density of the silicate-hydrogen phase would require a greater density deficit in metal to achieve neutral buoyancy. These are likely achievable, however, as metal density deficits greater than 0.55 (cf. Figure \ref{fig:terminal}) are possible at temperatures $>$ 5,000 K \citep{Schlichting_Young_2022}.   

We conclude that where reactions between the atmosphere and the core can provide a negative feedback between temperature and the density of the metal, there will be a point of neutral buoyancy that limits settling of metal to form a central metal core unless turbulence is lower than expected in the supercrtical magma oceans, or droplet sizes exceed hundreds of cm.

\begin{figure*}
   \centering
   \includegraphics[width=0.75\textwidth]{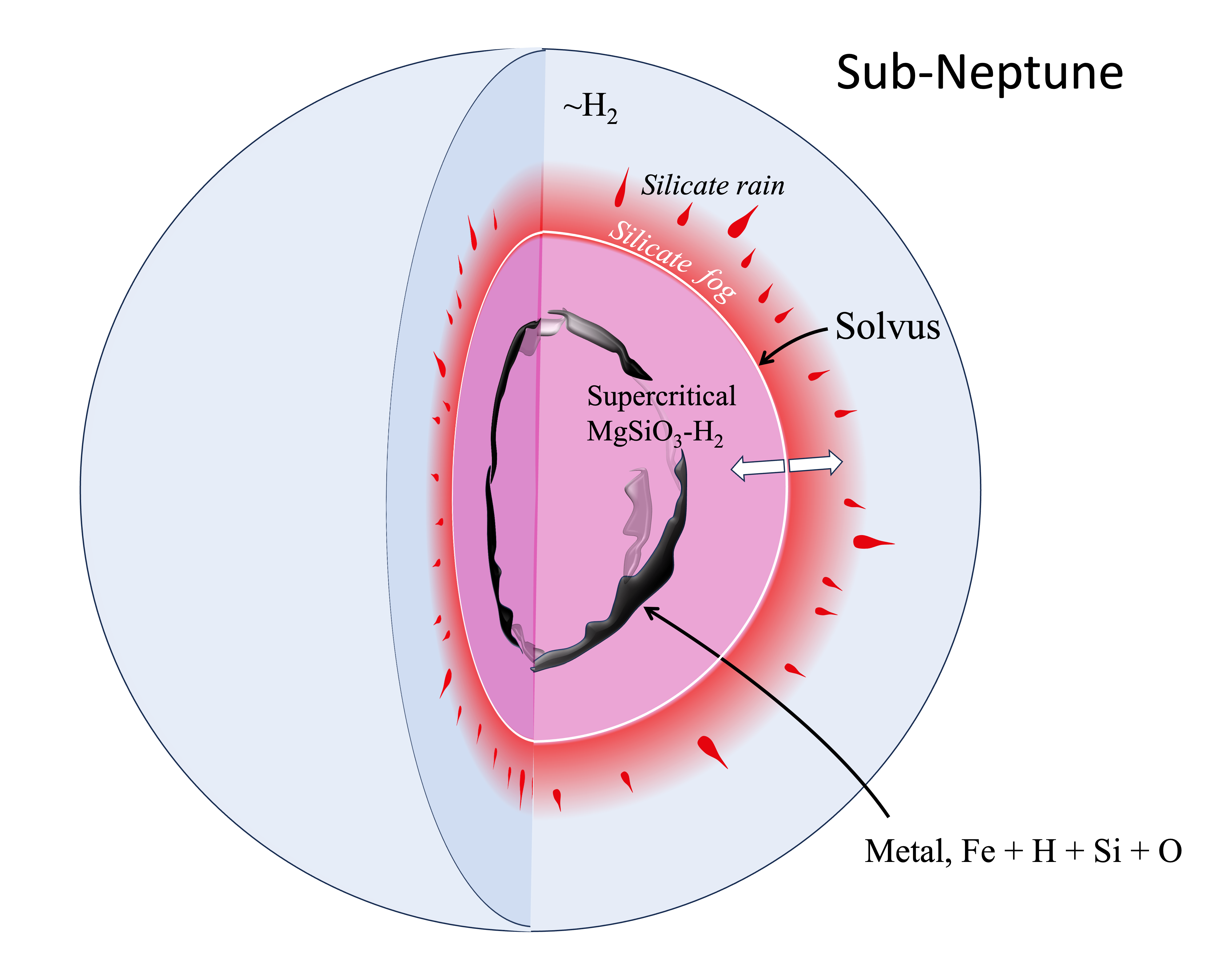}
    \caption{Interpretive diagram representing the consequences of the silicate-hydrogen solvus on the structure of sub-Neptunes. Moving inward from the outermost regions, the planet is composed of layers of nearly pure H$_2$, a layer of silicate precipition in a silicate-rich atmosphere, a layer of silicate fog in a high-molecular-weight gas phase just above the surface defined by the silicate-hydrogen solvus, and a supercritical silicate-hydrogen magma ocean where temperatures are greater than the solvus. The position of the solvus depends on the thermal state of the planet.  Under-dense metal segregations trapped near points of neutral buoyancy are shown in lieu of a central metal core.  }
\label{fig:planet}
\end{figure*}

\section{IMPLICATIONS FOR THE PHYSICAL AND CHEMICAL STRUCTURES OF PLANETS }
\label{section:structure}
 
Consideration of silicate-hydrogen miscibility provides  insights into the structure and chemistry of silicate-metal-hydrogen planets.  Figure \ref{fig:planet} shows a schematic of the structure of a sub-Neptune planet where the interior is still sufficiently hot to be supercritical.  The location of the silicate-hydrogen solvus defines the surface of the magma ocean, and depends on the total entropy of the system, which in turn depends on the age of the planet. Immediately above the solvus, one predicts there should be a convectively inhibited layer in which silicate ``fog" is present in the atmosphere if turbulence is sufficiently vigorous.  Higher in the atmosphere, droplets of condensate may grow sufficiently large that silicate precipitation is possible.  The silicate rain drops eventually sink at break-up velocities deeper in the atmosphere, feeding the layer of suspended droplets above the magma ocean. The longevity of the AMOI corresponding to the solvus in principle keeps the system in a steady state of silicate saturation.  This structure makes the boundary between the magma ocean and the overlying atmosphere less sharp.  

The picture in Figure \ref{fig:planet} is based on our fiducial 4$M_\oplus$ planet with a bulk hydrogen concentration of 4\% by mass, with much of the hydrogen in the supercritical core.  Planets formed with less hydrogen, to the low-H$_2$ side of the crest of the solvus (e.g., Figure \ref{fig:solvus1percent}), and those formed with greater concentrations of H$_2$ to the high side of the crest of the solvus (e.g., Figure \ref{fig:solvus10percent}) will have silicate with less hydrogen, and vapor phases with less silicate as a result of encountering the solvus at lower temperatures.  This is because at lower temperatures, the melt and gas limbs of the solvus are closer to the pure silicate and hydrogen-rich compositions, respectively. In the case of bulk hydrogen concentrations less than the solvus crest, atmospheres have nearly pure silicate condensates and only small fractions of gas, so the mass fraction of silicate condensate approaches unity throughout the atmosphere.  Conversely, for bulk compositions with greater H$_2$ than the crest of the solvus, the mass fraction of condensate is less than that shown in Figure \ref{fig:T_vs_R_4percent}D.   

Metal stranded by a neutral density contrast is shown schematically in Figure \ref{fig:planet}. Extensive miscibility of silicate and hydrogen may be yet another reason to consider that sub-Neptunes may not have well-defined central cores of metal \citep{Lichtenberg2021a, Elkins-Tanton_2008}.  

The predictions for mass fractions of primary atmospheres by \cite{ginzburg2016a} can be used to show that where conditions are suitable for a 4$M_\oplus$ planet to accrue  4\% H$_2$ by mass, a 1$M_\oplus$ planet would acquire about 1\% by mass H$_2$ from the protoplanetary disk. From calculations like these, one anticipates that planets to the low-H side of the solvus will include Earth-like bodies prior to loss of their primary atmospheres and the sub-Neptune progenitors of lower-mass super-Earths with core masses of $< 2M_\oplus$. Planets with bulk compositions to the high-H side of the solvus will include the progenitors of higher-mass super-Earths, sub-Neptunes and, where more volatiles are involved, Neptunes, with core masses $\ge 2M_\oplus$.

\section{Conclusions}
\label{section:conclusions}
An assessment of the pressures and temperatures associated with rocky planets with dense H$_2$-rich atmospheres suggests that a solvus depicting the miscibility of silicate and hydrogen  will dictate the structure of many sub-Neptune and their super-Earth descendants.  Temperatures and pressures at the base of the primary atmospheres will be controlled by the solvus at the hydrogen concentration of the bulk system.   In general, the maximum temperatures experienced at the surfaces of magma oceans should be less than predicted in the absence of silicate-hydrogen miscibility. The structures of the atmospheres differ from those calculated by omitting the miscibility of silicate and hydrogen.

The schematic applications of the silicate-hydrogen solvus presented here illustrate the rich chemical and physical impacts of even the simplest representation of the non-ideality in hydrogen-silicate mixing.  They also provide an \textit{a priori} means of estimating the conditions at the surfaces of magma oceans.

Based on this analysis, many sub-Neptunes are not expected to exhibit an Earth-like layering with discrete atmosphere, silicate mantle, and iron-rich core.  Indeed, it is not clear that  metal cores will form due to the high solubilities of H, Si, and O in iron metal at relevant conditions.   

In future, there remains a need for phase equilibria data in the system Mg-Si-Fe-O-H at conditions of approximately 3000 to 10000 K and 1 to 50 GPa, and higher $T$ and $P$ for metal-silicate equilibrium conditions.  While such conditions can be accessed in dynamic, laser-driven shock experiments where rapid, real-time characterization is required \citep[e.g.,][]{Fei2021,Alabdulkarim2022}, they are generally beyond those accessible for phase equilibria studies in which posterior characterization of products can be done  \citep[e.g.,][]{Baron2019}.  Ab initio calculations provide flexibility in both the conditions and the chemical systems to be interrogated, especially where machine learning can be used to accelerate coverage of composition space.

Also in future,  self-consistent models for planet formation that account for the role of the silicate-hydrogen solvus is required.  This is work in progress. Such a model is likely to modify our understanding of the structures of sub-Neptunes and their evolved super-Earth descendants.

\section{Acknowledgements} 
E.D.Y. acknowledges support from the AEThER program funded by the Alfred P. Sloan Foundation grant G202114194. E.D.Y. and L.S. acknowledge financial support from NASA grant 80NSSC24K0544 (Emerging Worlds program).  H.E.S. gratefully acknowledges NASA grant 80NSSC18K0828 for financial support during preparation and submission of the work.

\section*{Appendix}

We determine the structure of the core by solving Equations~\ref{eq:dmdr} through \ref{eq:dTdr} together with a specification of the thermodynamic properties of liquid silicate and liquid iron.  We use the fundamental thermodynamic relation (or thermodynamic potential) approach of \cite{DeKoker2009}.  All thermodynamic properties are determined by an analytical representation of the Helmholtz free energy as a function of volume and temperature, $F(V,T)$, that is fit to the results of ab initio molecular dynamics simulations.  The analytical form is inspired by Eulerian finite strain theory \citep{Birch1978} and the universal behavior of simple liquids \citep{Rosenfeld_Tarazona1998}.  Details of the functionoal form are given in \cite{DeKoker2009}.
The required material properties are then computed using analytical derivatives, including
\begin{equation}
    \rho=\frac{\mu}{V},
\end{equation}
\begin{equation}
    P=-\left( \frac{\partial F}{\partial V} \right)_T,
\end{equation}
\begin{equation}
    K_S= K_T \left(1 + \alpha \Gamma T \right),
\end{equation}
\begin{equation}
    \Gamma=\frac{\alpha K_T}{\rho c},
\end{equation}
\begin{equation}
    c = -\frac{T}{\mu} \left( \frac{\partial^2 F}{\partial T^2} \right)_V,
\end{equation}
with
\begin{equation}
    K_T= V \left( \frac{\partial^2 F}{\partial V^2} \right)_T,
\end{equation}
and,
\begin{equation}
    \alpha = -\frac{1}{K_T} \left( \frac{\partial^2 F}{\partial V\partial T} \right)_V, 
\end{equation}
where $K_T$ is the isothermal bulk modulus, and $\alpha$ is the thermal expansivity.

For liquid silicate, we use the fit to MgSiO$_3$ liquid properties previously detemined by \cite{DeKoker2009}, with an augmented molecular weight to account for the presence of iron in the rocky portion of planets ($\mu=20.34$ g/mol/atom).  For liquid iron, we fit to the ab initio molecular dynamics results of \cite{Wagle2019}, and adjust the molecular weight to match the density of Earth's core ($\mu$=52 g/mol) in order to allow for the effects of light elements in the metal.

We estimate the liquidus temperature of the mantle, $T_L$, as that for MgSiO$_3$ \citep{Deng2023}, where 
\begin{equation}
    T_L(\text{K})= \begin{cases}
        2875 \left( \frac{P-20}{8.11 \text{ GPa}}+1 \right)^{\frac{1}{3.73}} & P<172 \text{ GPa} \\
        5600 \left( \frac{P-120}{113.60 \text{ GPa}}+1 \right)^{\frac{1}{2.85}} & P>172 \text{ GPa},
    \end{cases}
    \label{eq_liquidus}
\end{equation}
corresponding to melting of the bridgmanite (lower pressure) and post-perovskite (higher pressure) phases.  We estimate the liquidus temperature of the metal core, where there is one, following \cite{Stixrude2014}:
\begin{equation}
    T_L(\text{K})=6500 \left(\frac{P}{340 \text{ GPa}} \right)^{0.515}(1-\ln x_0)^{-1}
\end{equation}
with the compostional variable $x_0=0.79$.

%\bibliography{edy_references}
\bibliographystyle{aasjournal}

%\end{linenumbers}
\end{document}